\documentclass[12pt,halfline,a4paper]{ouparticle}

\usepackage{natbib} 

\usepackage{makecell}

\usepackage{microtype}
\usepackage{graphicx}
\usepackage{subcaption}
\usepackage{booktabs} % for professional tables
\usepackage[table]{xcolor} % for rowcolor and color in tables

\usepackage{graphicx}
\usepackage{amsmath}
\usepackage{amssymb}
\usepackage{mathtools}
\usepackage{amsthm}

\usepackage{hyperref}
\usepackage[capitalize,noabbrev]{cleveref}

\usepackage{algpseudocode}
\usepackage{algorithmicx}

\usepackage{algorithm}        % provides the 

\theoremstyle{plain}
\newtheorem{theorem}{Theorem}[section]

\theoremstyle{definition}

\theoremstyle{remark}

\usepackage[utf8]{inputenc} % allow utf-8 input
\usepackage[T1]{fontenc}    % use 8-bit T1 fonts
\usepackage{hyperref}       % hyperlinks
\usepackage{url}            % simple URL typesetting
\usepackage{booktabs}       % professional-quality tables
\usepackage{amsfonts}       % blackboard math symbols
\usepackage{nicefrac}       % compact symbols for 1/2, etc.
\usepackage{microtype}      % microtypography
\usepackage{xcolor}         % colors

\usepackage{booktabs}
\usepackage{siunitx}
\usepackage[table]{xcolor}
\usepackage{caption}
\usepackage{subcaption}
\usepackage{tabularx}
\usepackage[utf8]{inputenc}
\usepackage[T1]{fontenc}

\usepackage{multirow}            % For cells spanning multiple rows
\usepackage{colortbl}            % For cell and row coloring
\usepackage{xcolor}              % Enhanced color support
\usepackage{tikz}                % For drawing arrows and custom elements
\usepackage{threeparttable}      % For tables with notes
\usepackage{array}               % Enhanced array and tabular environments
\usepackage{caption}             % For customizing captions
\usepackage{varwidth}            % For variable-width columns
\usepackage{bm}  

\usepackage{authblk}
\usepackage{lmodern}
\definecolor{navyblue}{RGB}{0,67,138}
\definecolor{ForestGreen}{RGB}{34,139,34}
\definecolor{BrickRed}{RGB}{178,34,34}
\definecolor{OliveGreen}{RGB}{85,107,47}
\definecolor{powderblue}{RGB}{176,224,230}
\definecolor{lime}{RGB}{191,255,0}

\begin{document}

\title{CodonMoE: DNA Language Models for mRNA Analyses}

\author{Shiyi Du, Litian Liang, Jiayi Li, and Carl Kingsford\thanks{Corresponding author.}}
\affil{\text{Ray and Stephanie Lane Computational Biology Department,}\\
\text{School of Computer Science,}\\
\text{Carnegie Mellon University, 5000 Forbes Avenue, Pittsburgh, PA}\\
\texttt{\{shiyid, litianl, lijiayi, carlk\}@cs.cmu.edu}}

\abstract{Genomic language models (gLMs) face a fundamental efficiency challenge: either maintain separate specialized models for each biological modality (DNA and RNA) or develop large multi-modal architectures. Both approaches impose significant computational burdens—modality-specific models require redundant infrastructure despite inherent biological connections, while multi-modal architectures demand massive parameter counts and extensive cross-modality pretraining. To address this limitation, we introduce CodonMoE (Adaptive Mixture of Codon Reformative Experts), a lightweight adapter that transforms DNA language models into effective RNA analyzers without RNA-specific pretraining. Our theoretical analysis establishes CodonMoE as a universal approximator at the codon level, capable of mapping arbitrary functions from codon sequences to RNA properties given sufficient expert capacity. Across four RNA prediction tasks spanning stability, expression, and regulation, DNA models augmented with CodonMoE significantly outperform their unmodified counterparts, with HyenaDNA+CodonMoE series achieving state-of-the-art results using 80\% fewer parameters than specialized RNA models. By maintaining sub-quadratic complexity while achieving superior performance, our approach provides a principled path toward unifying genomic language modeling, leveraging more abundant DNA data and reducing computational overhead while preserving modality-specific performance advantages.}

\date{}

\maketitle

\section{Introduction}

Recent advancements in  Large Language Models (LLMs) are revolutionizing  scientific disciplines, with the biomedical sciences experiencing especially profound effects~\citep{jumper2021highly,varadi2022alphafold}. The fundamental goal of Natural Language Processing (NLP) is to comprehend and manipulate sequences of words, a task that bears similarities to one of the central objectives in biology: deciphering the meaning and function encoded in biological sequences~\citep{eraslan2019deep}, as well generating novel genomic sequences with desired properties. This parallel has given rise to a new frontier in computational biology: Genomic Language Models (gLMs). GLMs are large-scale language models trained on vast amounts of biological sequence data. These models aim to capture the complex patterns and dependencies within genomic sequences, much like how general LLMs learn the intricacies of human language~\citep{bepler2021learning}. By leveraging the power of large language models and the abundance of genomic data now available, gLMs have the potential to significantly advance our understanding of genomes and reveal how DNA or RNA elements at various scales interact to give rise to biological functions~\citep{zhou2018deep}. 

Recent progress in state-space models (SSMs) have addressed the quadratic scaling limitations inherent in self-attention mechanisms, offering efficient alternatives to transformers for gLMs~\citep{ji2021dnabert,benegas2023gpn,ratcliff2024transformer} with subquadratic or linear scaling in sequence length. HyenaDNA~\citep{nguyen2024hyenadna}, built on the Hyena Hierarchy, represents a significant leap forward in genomic modeling, processing input contexts up to 1 million nucleotides --- a 500-fold increase over previous dense attention-based models. This architecture enables single-nucleotide-level analysis across extensive genomic regions, crucial for capturing long-range interactions and subtle genetic variations like SNPs. Caduceus~\citep{schiff2024caduceus}, leveraging the Mamba-based SSM~\citep{gu2023mamba}, introduces bi-directionality and reverse complementarity (RC) equivariance, essential properties for comprehensive DNA sequence analysis. Trained on $131$~kb sequences, Caduceus demonstrates superior performance on prediction of long-range effects of variants compared to much larger models. Building upon this framework, PlantCaduceus~\citep{zhai2024cross} extends these capabilities to diverse plant genomes, showcasing high transferability across species that diverged 160 million years ago and enabling genome-wide deleterious mutation identification without multiple sequence alignment. HELM \citep{yazdanijahromi2025helmhierarchicalencodingmrna} introduces a hierarchical encoding approach that integrates the biological codon structure of mRNA into language model training, acknowledging that multiple synonymous codons can encode identical amino acids while possessing distinct properties. EVO~\citep{nguyen2024sequence}, a hybrid architecture combining Hyena and Transformer elements, pushes the boundaries further with its 7 billion parameter model and 131 kb context length. EVO's multi-modal approach allows it to generalize across DNA, RNA, and protein prediction tasks, while also demonstrating unprecedented capabilities in generating synthetic molecular complexes and coding-rich sequences up to 650 kb in length.

The development of distinct gLMs --- encompassing DNA models, RNA models, and multimodal models --- introduces a considerable cost burden. This issue becomes increasingly pronounced as the size and complexity of gLMs grow. Moreover, attention-based models, particularly in the context of RNA language modeling, continue to dominate most RNA-specific tasks. Although these models deliver strong performance, their high computational demands remain a substantial challenge. DNA serves as the primary repository of genetic information, while mRNA functions as an intermediary in the expression of this information~\citep{crick1970central}. Building upon this fundamental concept, DNA-based language models offer a more foundational approach to genomic modeling compared to mRNA-based models. However, despite their great potential, DNA-based models have largely been underused for mRNA tasks. 

\begin{figure*}[htbp]
    \centering
    \includegraphics[width=0.9\textwidth]{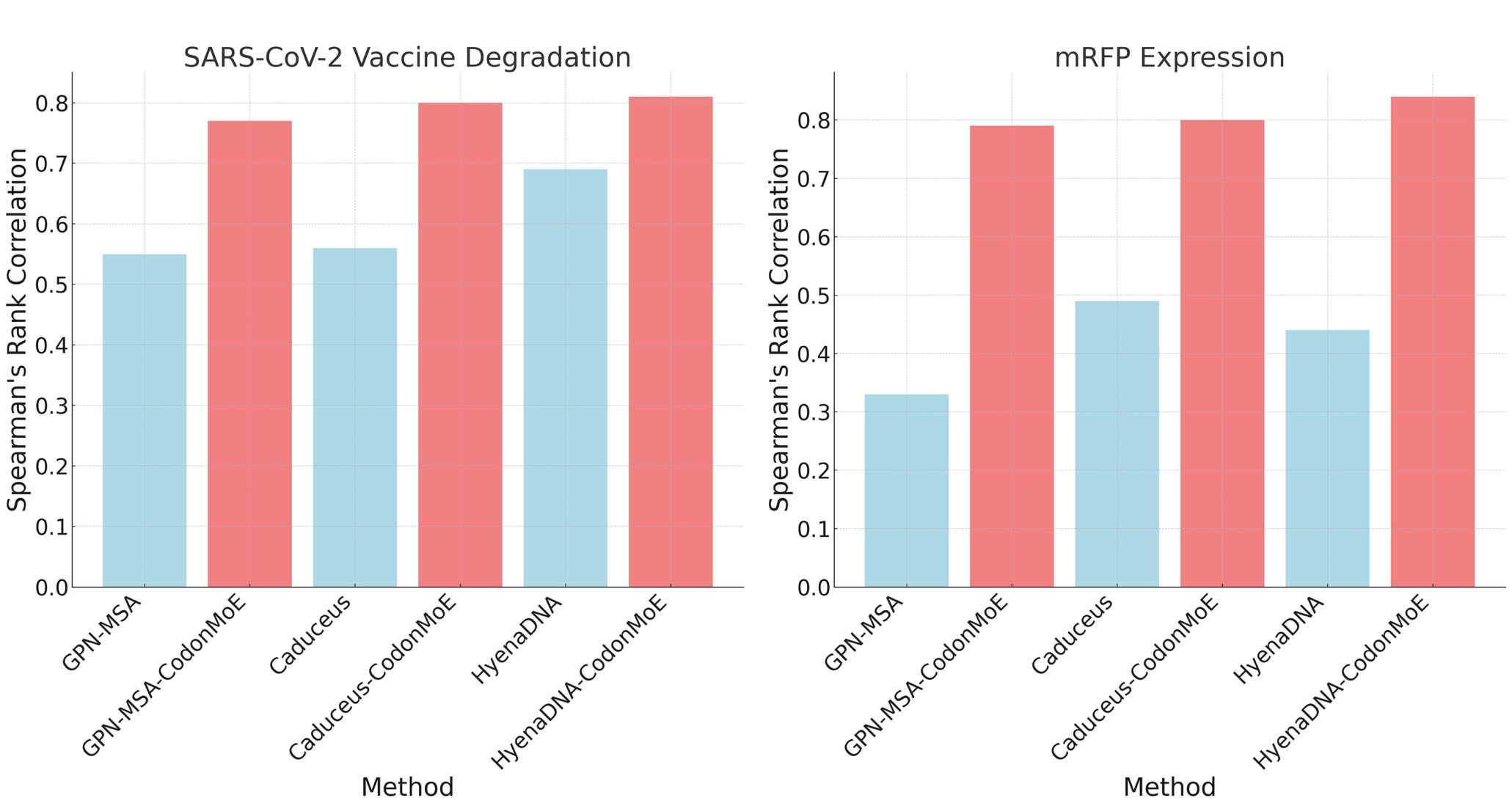}
    \caption{Performance comparison on prediction of mRFP expression and SARS-CoV-2 vaccine degradation across GPN-MSA~\citep{benegas2023gpn}, HyenaDNA~\citep{nguyen2024hyenadna}, and Caduceus~\citep{schiff2024caduceus} models, with (in light coral)  and without (in light blue) our CodonMoE integration.}
    \label{fig:performance_charts}
\end{figure*}

To address these challenges, we propose a novel approach based on the hypothesis that DNA models can effectively replace RNA models when augmented with RNA-specific control information. Central to our method is \textbf{Adaptive Mixture of Codon Reformative Experts (CodonMoE)}, a versatile plug-and-play module designed to seamlessly integrate with existing DNA models, transforming them into robust tools for mRNA analyses. We also demonstrate that recent, efficient sub-quadratic DNA-based state space model (SSM) architectures can be effectively combined with the CodonMoE to yield parameter- and computationally-efficient predictions for mRNA tasks. This marks the first approach to bridge the gap between DNA and RNA language models through a universally applicable adapter --- CodonMoE. Theoretical proof demonstrates that CodonMoE is a universal approximator of mRNA properties at the codon level. Experimental results further show that CodonMoE significantly enhances various DNA-based backbones by a wide margin, as illustrated in Figure~\ref{fig:performance_charts}. Some of these models achieve performance comparable to or exceeding state-of-the-art (SOTA) mRNA-specific models across several tasks, while also achieving substantial reductions in time complexity and model parameters.

To summarize our contributions:

\begin{itemize}
    \item  We introduce CodonMoE, a plug-and-play adapter that transforms DNA language models into effective RNA analyzers without requiring RNA-specific pretraining.
    
    \item We establish CodonMoE as a universal approximator at the codon level, proving it can map arbitrary functions from codon sequences to RNA properties given sufficient expert capacity.
    
    \item  Our approach maintains low complexity while achieving superior performance, with HyenaDNA+CodonMoE series achieving state-of-the-art results using 80\% fewer parameters than specialized RNA models.
    
    \item  Across RNA prediction tasks spanning stability, expression, and regulation, DNA models augmented with CodonMoE significantly outperform their unmodified counterparts.
    
    \item  By enabling DNA models to effectively handle RNA tasks, CodonMoE provides a principled path toward unifying genomic language modeling, leveraging more abundant DNA data while reducing the computational overhead of maintaining separate modality-specific models.
\end{itemize}

Source code for this work is available at \href{https://github.com/Kingsford-Group/CodonMoE}{https://github.com/Kingsford-Group/CodonMoE}.

\section{Related Work}
\label{gen_inst}

\paragraph{Transformer-based genomic language models} Transformers \citep{vaswani2017attention,devlin2018bert} have become prevalent in genomics modeling due to their capacity to capture long-range dependencies \citep{benegas2024genomic}. These models face challenges with extended sequences and nucleotide-level resolution. In DNA modeling, DNABERT~\citep{ji2021dnabert} employs k-mer tokenization for tasks like transcription factor binding site prediction. Enformer~\citep{avsec2021effective} incorporates convolution layers surrounding transformer blocks. Nucleotide Transformer \citep{dalla2023nucleotide} scales to five times DNABERT's size, while MegaDNA~\citep{shao2023long} extends context windows for longer sequences. GPN-MSA \citep{benegas2023gpn} uniquely leverages whole-genome alignments across species to enhance DNA modeling. For RNA applications, models such as RNABERT \citep{akiyama2022informative}, BigRNA~\citep{celaj2023rna}, CodonBERT \citep{li2024codonbert}, SpliceBERT \citep{chen2023self}, NicheFormer~\citep{schaar2024nicheformer} and scBERT \citep{yang2022scbert} address specialized transcriptomic tasks. These RNA-specific models encounter similar limitations as their DNA counterparts regarding sequence length and computational efficiency, motivating the exploration of state-space models as alternatives.

\paragraph{SSM-based genomic language models} State-space models have gained prominence in genomic modeling by offering reduced computational complexity and improved scaling for long-range dependencies. HyenaDNA \citep{nguyen2024hyenadna} and Caduceus \citep{schiff2024caduceus} demonstrate effectiveness in sequence modeling with SSM architectures. EVO \citep{nguyen2024sequence} further extends these capabilities to whole-genome-scale DNA generation, while also demonstrating cross-modality applicability to RNA. HELM \citep{yazdanijahromi2025helmhierarchicalencodingmrna} introduces a hierarchical encoding approach that explicitly accounts for the codon structure of mRNA, where multiple synonymous codons can encode the same amino acid, highlighting the importance of incorporating biological structure into model training.

\paragraph{Mixture of Experts} MoE architectures enhance model performance through specialized experts focusing on different input aspects. Initially proposed by \citet{jacobs1991adaptive} and extended hierarchically by \citet{jordan1994hierarchical}, MoE was adapted for NLP by \citet{shazeer2017outrageously} through sparsely-gated activation. Subsequent developments include GShard's scaling beyond 600 billion parameters \citep{lepikhin2021gshard}, Switch Transformer's single-expert routing \citep{fedus2021switch}, GLaM's energy-efficient scaling \citep{du2021glam}, and improved sparsity management \citep{zuo2022taming}. Modern MoE implementations enable efficient scaling while reducing computational requirements through dynamic expert activation.

\section{Methodology}

\begin{figure*}[h]
\centering
\includegraphics[width=1.0 \textwidth ]{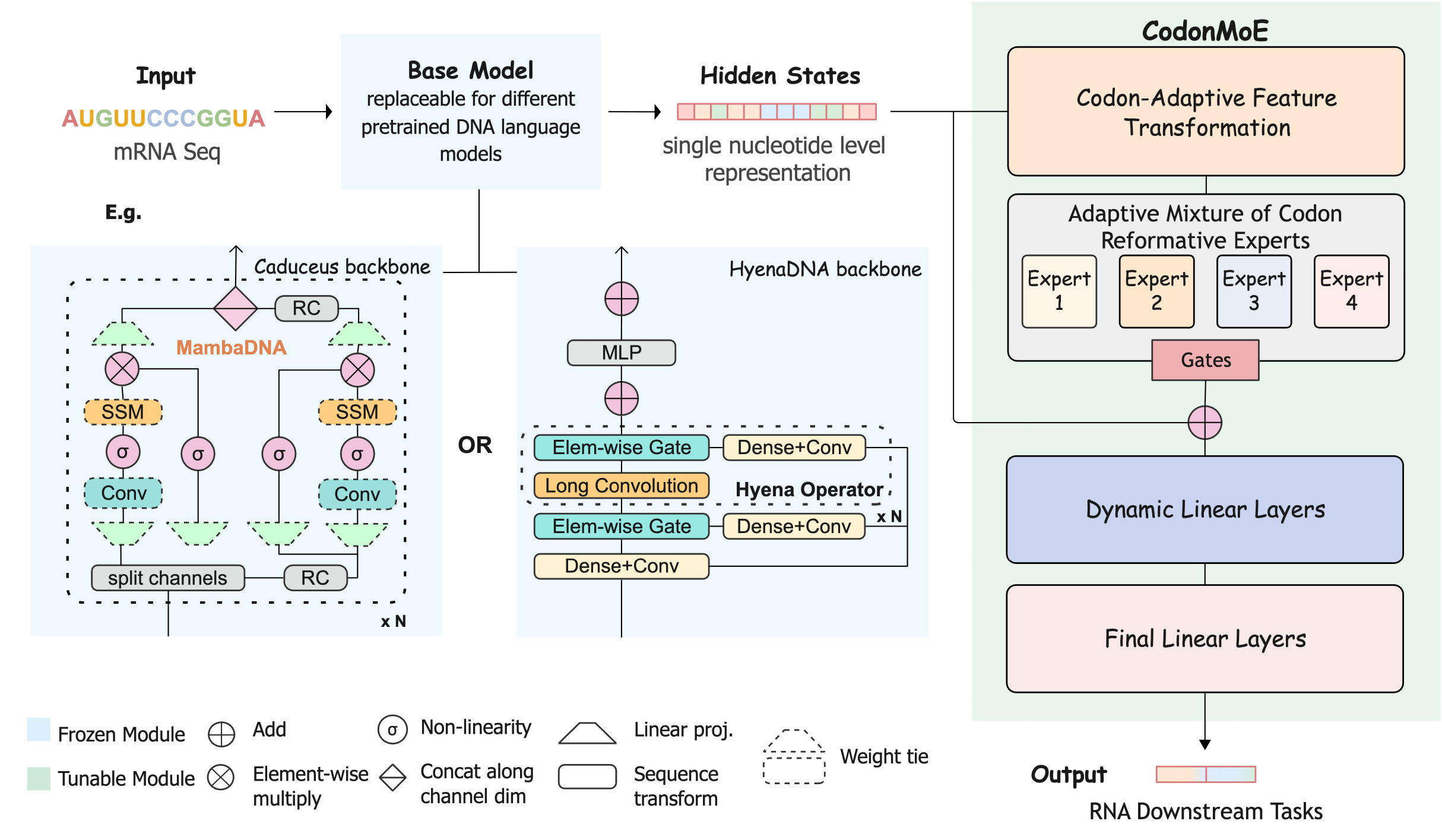}
\caption{Overview of CodonMoE and proposed framework. The architecture combines pretrained DNA-focused state space models with a novel CodonMoE module. This CodonMoE adapts DNA-derived patterns for RNA analysis by grouping inputs into codons and using a Mixture of Experts (MoE) approach. This design enables effective translation of DNA models for RNA sequence analysis, leveraging the strengths of both domains. } 
\label{fig:overview}
\end{figure*}

We introduce a novel module, CodonMoE, that can be integrated into state-of-the-art pretrained SSMs and attention-based models designed for DNA sequence analysis to adapt them for RNA analyses (Figure~\ref{fig:overview}). The CodonMoE processes these hidden states from those DNA backbones by restructuring the input into codons (three-nucleotide sequences) and applying an Adaptive Mixture of Codon Reformative Experts. Each expert within the CodonMoE is designed to identify and emphasize various biological signals, enabling the model to capture both codon-level and broader sequence patterns. Furthermore, we demonstrate that the CodonMoE is a universal approximator at the codon level. Given sufficient expert capacity, the CodonMoE can approximate any function that maps codon sequences to specific target properties with arbitrary precision when combined with the pretrained backbone model.  In general, this architecture effectively translates DNA models to RNA contexts, allowing for robust analysis of RNA sequences.
\label{back}

\subsection{CodonMoE: adaptive mixture of codon reformative experts}

\paragraph{Sample-wise dynamic codon-level representation} 
CodonMoE processes representations of codons, which are groups of three nucleotides in genetic sequences encoding amino acids. The input to CodonMoE consists of nucleotide representations with dynamic dimensionality, allowing it to accommodate input samples of varying sequence lengths. These inputs are reshaped into codon groups, preserving the structure of the genetic code. The CodonMoE slices this sequence to extract codon-related segments and reshapes them to facilitate further processing.

\paragraph{Adaptive mixture of reformative codon experts}
CodonMoE uses Adaptive Mixture of Codon Experts layers, where $k$ experts, each specializing in different aspects of the codon data, process these representations. The transformation is given by:
\[
y_{\text{codons}}^{\text{MoE}} = \sum_{k=1}^{K} g_k(x)\cdot E_k(y_{\text{codons}}),
\]
where \(g_k(x)\) is the gating mechanism that determines the contribution of each expert \(E_k\). This dynamic expert selection allows the MoE to process the codon data in multiple ways, with the gating system controlling which perspective should dominate.

\paragraph{Dynamic reshaping and contextualization} After processing by the experts, the codon-level representations are reshaped to match the original sequence length and structure. The CodonMoE contextualizes this information, enriching it with surrounding data before recombining it with the rest of the input sequence:
\[
y_{\text{output}} = y_{\text{reshaped}} + y_{\text{codons}}^{\text{MoE}}.
\]
This process ensures that codon-level information is properly embedded and aligned within the original sequence, helping the model recognize both local codon-specific patterns and broader genetic patterns.
For more detailed specification of the algorithm, refer to the
appendix~\ref{algorithmpseudocode}.

Building on CodonMoE, CodonMoE-pro replaces the final layers with a codon neighborhood convolution. The convolutional mechanism in CodonMoE-pro processes these distilled features generated from CodonMoE gates. By applying codon neighborhood convolutional operations to these specialized representations, the higher-order relationships between the already-refined codon features can be detected. This hierarchical processing pipeline enhances the model's ability to capture biologically relevant patterns that influence translation dynamics. The architecture leverages the biological insight that codon context, rather than isolated codons, often determines translation efficiency and mRNA half-life, thereby providing a more nuanced approach to sequence analysis that aligns with current understanding of translation biology.

\subsection{CodonMoE is a universal approximator at the codon level}

We show that given sufficient capacity, our proposed CodonMoE can approximate any function that maps codon sequences to target properties with arbitrary precision when integrated with the pretrained backbone model.

\textbf{DNA sequence space} ($\mathcal{X}$) is defined as the set of all possible DNA sequences composed of nucleotides from the alphabet $\{A, C, G, T\}$.  In our framework, the RNA nucleotide `$U$' is replaced with the DNA nucleotide `$T$', aligning RNA codons with their corresponding DNA representations. \textbf{Codon Space} ($\mathcal{C}$) consists of all possible codons, where each codon is a sequence of three nucleotides from $\{A, C, G, T\}$. Formally, $\mathcal{C} = \{A, C, G, T\}^3$. \textbf{Function Class} ($\mathcal{F}$) comprises all continuous functions $f: \mathcal{C}^n \rightarrow \mathbb{R}$ that map sequences of $n$ codons to specific target properties.

Our modeling approach is structured around a two-stage paradigm. Initially, a \textbf{Backbone Model} $h: \mathcal{X} \rightarrow \mathbb{R}^{L \times D}$ is pretrained on DNA sequences, where $L$ represents the sequence length and $D$ the embedding dimension. This pretraining phase equips the backbone with foundational knowledge of genetic sequences and their inherent patterns. We directly use the pretrained models on DNA
sequences. Subsequently, the \textbf{CodonMoE} serves as an adapter to this pretrained backbone model. Formally, the CodonMoE is a function $g: \mathbb{R}^{L \times D} \rightarrow \mathbb{R}$ that is fine-tuned on mRNA sequences to specialize the model for mRNA-specific tasks. This fine-tuning process involves training the adapter using mRNA sequences.

\begin{theorem}
\label{thm:universal_approximation}
Let $\mathcal{C} = \{A, C, G, T\}^3$ be the codon space, and let $\mathcal{F} = \{f: \mathcal{C}^n \rightarrow \mathbb{R}\}$ be the class of target functions. Consider a pretrained backbone model $h: \mathcal{X} \rightarrow \mathbb{R}^{L \times D}$, where $\mathcal{X} = \{A, C, G, T\}^*$, and an adapter CodonMoE $g: \mathbb{R}^{L \times D} \rightarrow \mathbb{R}$ structured as a dense MoE with $K$ experts. Assume the following conditions hold:
\begin{enumerate}
    \item \textbf{Expert capacity}: Each expert $E_k: \mathbb{R}^{D} \rightarrow \mathbb{R}^{D'}$ within the MoE is a neural network capable of uniformly approximating any continuous function on compact subsets of $\mathbb{R}^{D}$.
    \item \textbf{Gating mechanism}: The gating network $G: \mathbb{R}^{D} \rightarrow \Delta^K$ (where $\Delta^K$ is the $K$-simplex) assigns non-negative weights $g_k(z_i)$ to each expert based on the input $z_i \in \mathbb{R}^{D}$, satisfying $\sum_{k=1}^K g_k(z_i) = 1$.
    \item \textbf{Embedding representation}: Each DNA sequence $x \in \mathcal{X}$ is partitioned into codons $(c_1, c_2, \dots, c_n)$, and the backbone model generates embeddings $h(x) \in \mathbb{R}^{L \times D}$, where $L = 3n$ (assuming each codon is represented by three consecutive embeddings).
\end{enumerate}
Then, for any function $f \in \mathcal{F}$ and for any $\epsilon > 0$, there exists a number of experts $K$ and corresponding parameters for the CodonMoE such that, for all $x \in \mathcal{C}^n$, the approximation error satisfies
\[
\left| f(c_1, c_2, \dots, c_n) - g\left( \sum_{i=1}^n \sum_{k=1}^K g_k(z_i) \cdot E_k(z_i) \right) \right| < \epsilon,
\]
where $z_i = [h(c_i)] \in \mathbb{R}^{D}$ is codon $c_i$ represented by averaging three nucleotide embeddings.
\end{theorem}

For proof of Theorem~\ref{thm:universal_approximation}, refer to the
appendix~\ref{proof}.

\section{Experiments}
\label{others}

\subsection{Tasks and datasets}

% !!more conceise version
We evaluate CodonMoE on $4$ RNA tasks, spanning different aspects of RNA biology. The monomeric Red Fluorescent Protein (mRFP) dataset by \citet{nieuwkoop2023revealing} contains $1{,}459$ mRFP variants with paired expression levels and sequences, derived from codon-randomized libraries with varying adaptation index biases to analyze sequence variation effects on expression. The task is to predict the expression of these variants from their sequence. The SARS-CoV-2 vaccine degradation dataset by \citet{leppek2022combinatorial} comprises $2{,}400$ samples with measurements of vaccine stability/degradation, providing insights into mRNA vaccine durability factors. The goal is to predict a sequence's stability. The Tc-riboswitch dataset \citep{groher2018tuning} contains $355$ tetracycline riboswitch dimer sequences positioned upstream of a GFP coding region, with the switching factor measuring differential regulatory effects in the presence versus absence of tetracycline. The goal is to predict the efficacy of the riboswitch. The MLOS dataset \citep{li2024codonbert} includes $164$ mRNA candidates encoding influenza hemagglutinin antigen, with fixed untranslated regions and variable coding regions, evaluated for protein expression levels in HeLa cells. The goal is to predict protein levels derived from the given sequence. All datasets were selected to evaluate model performance and match those used in HELM \citep{yazdanijahromi2025helmhierarchicalencodingmrna} or CodonBERT \citep{li2024codonbert}, facilitating direct comparison across key mRNA tasks using consistent 70\%/15\%/15\% train/validation/test splits. All experiments were completed on a single NVIDIA A100 GPU. For more details, see appendix section \ref{appendix:experimentalsettings}.

\subsection{Model configurations and baselines}

Our evaluation spans diverse backbone architectures to thoroughly assess the CodonMoE series (standard CodonMoE and parameter-optimized CodonMoE-pro) as cross-modality adapters. We implement the CodonMoE series across four distinct DNA model paradigms: linear complexity models (GPN-SS~\citep{benegas2023gpn}, Caduceus~\citep{schiff2024caduceus}), quadratic complexity attention-based models (GPN-MSA~\citep{benegas2023gpn}), and sub-quadratic non-attention models (HyenaDNA~\citep{nguyen2024hyenadna}). This  allows us to determine which architectural characteristics are most amenable to cross-modality adaptation and whether efficiency benefits transfer consistently across model families.

For rigorous benchmarking, we compare against two categories of baseline methods: (1) state-of-the-art RNA foundation models (CodonBERT~\citep{li2024codonbert}, RNA-FM~\citep{li2024codonbert,chen2022interpretable}, SpliceBERT~\citep{chen2023self, yazdanijahromi2025helmhierarchicalencodingmrna}, and various transformer/state-space architectures including Transformer XE, Hyena XE, Mamba XE, and Transformer HELM~\citep{yazdanijahromi2025helmhierarchicalencodingmrna,vaswani2017attention,nguyen2024hyenadna,gu2023mamba}), and (2) classical feature-engineering approaches (TF-IDF, TextCNN, Codon2vec~\citep{rajaraman2011mining, zhang2015sensitivity, li2024codonbert}). This comprehensive comparison framework establishes whether adapting efficient DNA models can outperform both traditional methods and specialized RNA-specific architectures while maintaining computational advantages.

Our analysis examines both absolute performance and relative improvements over baseline models, with special attention to the efficiency-performance trade-offs achieved through cross-modality adaptation using the CodonMoE series. We measure performance using Spearman's rank correlation, the field-standard metric for assessing biological property prediction independent of absolute scale. Implementation details including backbones, hyperparameters, training procedures, and optimization settings are documented in the appendix section~\ref{appendix:experimentalsettings} and~\ref{appendix: intro2backbone}.

\subsection{Results}
\label{results}
\paragraph{Unlocking DNA models for mRNA analyses}
\sloppy
Table~\ref{tab:dna_with_codonmoe} demonstrates our CodonMoE series consistently enhances DNA language models for RNA analysis tasks. HyenaDNA+CodonMoE-pro achieves state-of-the-art results on vaccine degradation and mRFP expression prediction with only 7.5M parameters and sub-quadratic complexity, outperforming specialized RNA models that require $10\times$ more parameters. Notably, models with initially poorer RNA task performance showed the most dramatic improvements, suggesting CodonMoE fundamentally bridges DNA-RNA representational gaps rather than providing incremental enhancements.

% Table 1\
\begin{table}[!htbp]
    \caption{Comparison of DNA language model backbones with and without our CodonMoE adapters. We report model modality, computational complexity, total parameters, and Spearman’s rank correlation on SARS-CoV-2 vaccine degradation and mRFP expression tasks. }
    \label{tab:dna_with_codonmoe}
    \centering
    \small
    \setlength{\tabcolsep}{2.5pt} 
    \begin{tabular}{lccccc}
    \toprule
    \rowcolor{gray!10} 
    \textbf{Method} & \textbf{Modality} & \textbf{Complexity} & \textbf{Parameters} & \textbf{Vaccine} & \textbf{mRFP} \\

    \midrule
    % \multicolumn{6}{c}{\textbf{DNA Models}} \\
    GPN-SS & DNA & $O(L)$ & 65.6M & 0.60 & 0.56 \\
    \rowcolor{blue!5} GPN-SS+CodonMoE (ours) & DNA & $O(L)$ & 78.2M & \cellcolor{lime!7}0.74 & \cellcolor{lime!15}0.82 \\
    \addlinespace
    GPN-MSA & DNA & $O(L^2)$ & 85.7M & 0.55 & 0.33 \\
    \rowcolor{blue!5} GPN-MSA+CodonMoE (ours) & DNA & $O(L^2)$ & 161.9M & \cellcolor{lime!7}0.77 & \cellcolor{lime!15}0.79 \\
    \rowcolor{blue!5} GPN-MSA+CodonMoE-pro (ours) & DNA & $O(L^2)$ & 115.0M & \cellcolor{lime!15}0.82 & \cellcolor{lime!15}0.81 \\
    \addlinespace
    Caduceus & DNA & $O(L)$ & 7.7M & 0.56 & 0.49 \\
    \rowcolor{blue!5} Caduceus+CodonMoE (ours) & DNA & $O(L)$ & 48.1M & \cellcolor{lime!15}0.80 & \cellcolor{lime!15}0.80 \\
    \addlinespace
    HyenaDNA & DNA & $O(L\log L)$ & 4.1M & 0.69 & 0.44 \\
    \rowcolor{blue!5} HyenaDNA+CodonMoE (ours) & DNA & $O(L\log L)$ & 12.7M & \cellcolor{lime!15}0.81 & \cellcolor{lime!20}0.84 \\
    \rowcolor{blue!10} HyenaDNA+CodonMoE-pro (ours) & DNA & $O(L\log L)$ & 7.5M & \cellcolor{lime!25}\textbf{0.84}$^*$ & \cellcolor{lime!25}\textbf{0.88}$^*$ \\
    \bottomrule
    \multicolumn{6}{p{0.98\linewidth}}{\footnotesize \textbf{Notes:} All models evaluated on identical train/validation/test splits (70\%/15\%/15\%), using the same split set as in the CodonBERT~\citep{li2024codonbert}. $^*$ = Best overall performance. Lime green background intensity indicates relative performance level.}\\
    \end{tabular}
\end{table}

\paragraph{Quantifying adaptation effectiveness}
Table~\ref{tab:performance_boost} quantifies CodonMoE's improvement across DNA backbones, with GPN-MSA showing the most substantial gains, particularly on mRFP expression. This pattern reflects how CodonMoE addresses the inherent limitations of different architectures. For instance, GPN-MSA exhibits a ``smoothing effect'' on rare codon signals that influence mRNA translation efficiency, which CodonMoE corrects by providing dedicated expert pathways for these rare but influential codon contexts.

% Table 2: Performance Boost Table
\begin{table}[H]
    \caption{Performance gain analysis: Quantifying improvements
    from CodonMoE integration across DNA language model backbones.}
    \label{tab:performance_boost}
    \centering
    % \small
    \footnotesize
    \setlength{\tabcolsep}{2.5pt} 
    \begin{tabular}{lccccccc}
    \toprule
    \rowcolor{gray!10}
    \textbf{DNA Backbone} 
    & \multicolumn{3}{c}{\textbf{Vaccine degradation}}
    & \multicolumn{3}{c}{\textbf{mRFP expression}}
    & \textbf{Avg} \\
    \cmidrule(lr){2-4}\cmidrule(lr){5-7}
    \rowcolor{gray!10}
    & \textbf{Base} & \textbf{+CodonMoE (ours)} & \textbf{$\Delta$}
    & \textbf{Base} & \textbf{+CodonMoE (ours)} & \textbf{$\Delta$}
    & \textbf{$\Delta$} \\
    \midrule
    GPN-SS    
    & 0.60 
    & 0.74 
    & 0.14\textcolor{ForestGreen}{$\uparrow$} 
    & 0.56 
    & 0.82 
    & 0.26\textcolor{ForestGreen}{$\uparrow\uparrow$} 
    & 0.20 \\
    
    \rowcolor{gray!3}
    GPN-MSA   
    & 0.55 
    & 0.77 
    & 0.22\textcolor{ForestGreen}{$\uparrow\uparrow$} 
    & 0.33 
    & 0.79 
    & 0.46\textcolor{ForestGreen}{$\uparrow\uparrow\uparrow$} 
    & 0.34 \\
    
    Caduceus  
    & 0.56 
    & 0.80 
    & 0.24\textcolor{ForestGreen}{$\uparrow\uparrow$} 
    & 0.49 
    & 0.80 
    & 0.31\textcolor{ForestGreen}{$\uparrow\uparrow$} 
    & 0.28 \\
    
    \rowcolor{gray!3}
    HyenaDNA  
    & 0.69 
    & 0.81 
    & 0.12\textcolor{ForestGreen}{$\uparrow$} 
    & 0.44 
    & 0.84 
    & 0.40\textcolor{ForestGreen}{$\uparrow\uparrow\uparrow$} 
    & 0.26 \\
    \bottomrule
    \multicolumn{8}{p{0.98\linewidth}}{\footnotesize \textbf{Notes:} $\Delta = (+\text{CodonMoE}) - (\text{Base})$ represents absolute improvement in Spearman's $\rho$. Green arrows indicate magnitude: $\uparrow$ (small), $\uparrow\uparrow$ (medium), $\uparrow\uparrow\uparrow$ (large).}\\
    \end{tabular}
\end{table}
    
\paragraph{Generalization across multiple mRNA tasks}
CodonMoE-pro consistently outperforms standard CodonMoE across representative backbones (Table~\ref{tab:performance_boost_pro_specific}), suggesting a ``threshold effect'' where 7--12M parameters provide optimal coverage of critical codon contexts without overfitting. Our best configuration, HyenaDNA+CodonMoE-pro, achieves state-of-the-art results on three out of four RNA tasks (Table~\ref{tab:model_comparison_extended}), demonstrating that properly adapted DNA models with minimal parameters can outperform specialized RNA models across diverse RNA functionality domains. For more comparison, see appendix section~\ref{appendix:comprehensive_comparison}.

% Table 3: CodonMoE-pro Table
\begin{table}[H]
    \caption{Comparative analysis: CodonMoE-pro adapter yields enhanced performance on DNA backbones.}
    \label{tab:performance_boost_pro_specific}
    \centering
    % \small
    \footnotesize
    \setlength{\tabcolsep}{2.5pt} 
    \begin{tabular}{lccccccc}
    \toprule
    \rowcolor{gray!10}
    \textbf{DNA Backbone} 
    & \multicolumn{3}{c}{\textbf{Vaccine degradation}}
    & \multicolumn{3}{c}{\textbf{mRFP expression}}
    & \textbf{Avg} \\
    \cmidrule(lr){2-4}\cmidrule(lr){5-7}
    \rowcolor{gray!10}
    & \textbf{Base} & \textbf{+CodonMoE-pro (ours)} & \textbf{$\Delta$}
    & \textbf{Base} & \textbf{+CodonMoE-pro (ours)} & \textbf{$\Delta$}
    & \textbf{$\Delta$} \\
    \midrule
    GPN-MSA   
    & 0.55 
    & 0.82
    & {0.27} 
    & 0.33 
    & 0.81 
    & {0.48} 
    & 0.38 \\
    \midrule
    HyenaDNA  
    & 0.69 
    & 0.84$^\ddagger$ 
    & {0.15} 
    & 0.44 
    & 0.88$^\ddagger$ 
    & 0.44 
    & 0.29 \\
    \bottomrule
    \multicolumn{8}{p{0.98\linewidth}}{\footnotesize \textbf{Notes:} $^\ddagger$ = New state-of-the-art results despite using DNA pretrained backbones. CodonMoE-pro consistently outperforms standard CodonMoE (compare with Table~\ref{tab:performance_boost}).}\\
    \end{tabular}
\end{table}

% Table: RNA/DNA Models Comparison with MLOS 
\begin{table}[H] 
    \caption{Evaluation of computational complexity and Spearman's rank correlation metrics across RNA and DNA language models: measuring the impact of CodonMoE-pro integration on model performance and parameter efficiency. The metric is Spearman's rank correlation.} 
    \label{tab:model_comparison_extended} 
    \centering 
    \footnotesize 
    \setlength{\tabcolsep}{2.5pt} 
    \begin{tabular}{lccccccc} 
    \toprule 
    \rowcolor{gray!10} 
    \textbf{Model} & \textbf{Params} & \textbf{Size} & \textbf{Complexity} & \textbf{Vaccine} & \textbf{mRFP} & \textbf{Tc-ribo.} & \textbf{MLOS} \\ 
    \midrule 
    \multicolumn{8}{l}{\textit{\textbf{RNA Foundation Models}}} \\ 
    RNA-FM & 100.00M & \textcolor{red}{$\square\square\square$} & $O(L^2)$ & 0.74 & 0.80 & 0.58$^\ddagger$ & -- \\ 
    SpliceBERT & 20.00M & \textcolor{orange}{$\square\square$} & $O(L^2)$ & 0.76 & 0.60 & 0.42 & -- \\ 
    CodonBERT & 81.70M & \textcolor{red}{$\square\square\square$} & $O(L^2)$ & 0.77 & 0.85$^\dagger$ & 0.56 & 0.54 \\ 
    Transformer XE & 50.00M & \textcolor{orange}{$\square\square$} & $O(L^2)$ & 0.79$^\ddagger$ & 0.82 & 0.53 & 0.61$^\ddagger$ \\ 
    Hyena XE & 50.00M & \textcolor{orange}{$\square\square$} & $O(L\log L)$ & 0.80$^\dagger$ & 0.84$^\ddagger$ & 0.52 & 0.62$^\dagger$ \\ 
    Mamba XE & 50.00M & \textcolor{orange}{$\square\square$} & $O(L)$ & 0.74 & 0.82 & 0.52 & 0.62$^\dagger$  \\ 
    Transformer HELM & 50.00M & \textcolor{orange}{$\square\square$} & $O(L^2)$ & 0.79$^\ddagger$ & 0.85$^\dagger$ & \textbf{0.62} & 0.59 \\ 
    \midrule 
    \multicolumn{8}{l}{\textit{\textbf{DNA Foundation Models Enhanced with CodonMoE}}} \\ 
    \rowcolor{blue!10} 
    HyenaDNA + CodonMoE-pro (ours) & 7.50M & \textcolor{green}{$\square$} & $O(L\log L)$ & \textbf{0.84} & \textbf{0.88} & 0.60$^\dagger$ & \textbf{0.63} \\ 
    \bottomrule 
    \end{tabular} 
    \vspace{1mm} 
    \begin{tablenotes} 
    \footnotesize 
    \item \textbf{Notes:} \textbf{Size:} \textcolor{green}{$\square$} < 10M, \textcolor{orange}{$\square\square$} = 20-80M, \textcolor{red}{$\square\square\square$} > 80M; \textbf{Complexity:} $O(L)$ - linear, $O(L\log L)$ - linearithmic, $O(L^2)$ - quadratic ($L$ = sequence length); \textbf{Performance:} \textbf{Bold} = best, $^\dagger$ = second-best, $^\ddagger$ = third-best; Missing values:
    model unable to process due to sequence length limitations; HyenaDNA-CodonMoE-pro achieves SOTA results with only 9\% of CodonBERT's parameters. 
    \end{tablenotes} 
 \end{table}

\subsection{Ablation studies}
\subsubsection{Comparative analysis of codon operator variants}

To systematically evaluate architectural components, we compare three variants of codon operators with increasing complexity: CodonMean (nucleotide averaging), CodonMoE (expert-based routing), and CodonMoE-pro (with codon neighborhood convolution). Table~\ref{table2_mrfp} demonstrate consistent performance gains across variants, with CodonMoE-pro achieving the highest Spearman's rank correlation on both mRFP expression and vaccine degradation tasks.

\begin{table}[htbp]
\setlength{\abovecaptionskip}{0cm}
\centering
\caption{Codon operator variant comparison on  mRFP expression and vaccine degradation (Spearman's rank correlation).}
\label{table2_mrfp}
\footnotesize

\begin{tabular}{lcccc}
\toprule
  & \multicolumn{2}{c}{\bf mRFP} & \multicolumn{2}{c}{\bf Vaccine degradation}\\
  \cmidrule(lr){2-3}\cmidrule(lr){4-5}
  &\bf GPN-MSA &\bf HyenaDNA &\bf GPN-MSA &\bf HyenaDNA  \\
\midrule
CodonMean  & 0.740  & 0.765 & 0.729 & 0.789 \\
CodonMoE  & 0.790 & 0.837 & 0.770 & 0.812 \\
CodonMoE-pro  & \textbf{0.808} & \textbf{0.878} & \textbf{0.823} & \textbf{0.844}\\
\bottomrule
\end{tabular}
\end{table}

\paragraph{Architectural benefits beyond parameter count}
Table~\ref{tab:dense_ablation} shows that HyenaDNA+CodonMoE-pro significantly outperforms a parameter-matched dense baseline (see Appendix~\ref{appendix：ablation_dense}), demonstrating that the performance gains derive from architectural design rather than simply increased capacity. The MoE architecture provides three key benefits: (1) optimized parameter arrangement that prioritizes structure over scale, (2) specialized handling of functionally critical codons, and (3) inherent regularization through sparse activation patterns.

Our ablations reveal distinct task-specific requirements: stability tasks benefit from global pattern recognition while expression tasks require fine-grained n-gram detection, explaining CodonMoE-pro's strong performance across diverse RNA functionality domains. For detailed mechanism analysis and additional experiments, refer to Appendix~\ref{appendix:visualization}, ~\ref{appendix：codonmoepro_effectiveness} and~\ref{appendix:detailed_ablation}.

\subsubsection{Raw DNA embeddings lack mRNA-specific guidance}
\label{appendix:raw_dna_embeddings}
To probe how much mRNA-relevant information DNA gLMs learn “out of the box,” we evaluated two representative backbones --- GPN-MSA (attention-based) and HyenaDNA (non-attention-based) --- with minimal adaptation (Table~\ref{tab:raw_embeddings_dna}). We froze their pretrained embeddings and trained two standard regressors, an MLP and an XGBoost model, on the mRFP expression and SARS-CoV-2 degradation tasks. Even with a powerful tree-based regressor, raw embeddings only reach limited performance on mRFP—well below RNA-trained models. This plateau shows DNA gLMs need explicit mRNA-specific guidance.

\begin{table}[htbp]
\setlength{\abovecaptionskip}{0cm}
\centering
\caption{Spearman’s rank correlation on mRFP expression and SARS- CoV- 2 vaccine degradation tasks using raw DNA model embeddings.}
\label{tab:raw_embeddings_dna}
\footnotesize
\begin{tabular}{lcccc}
\toprule
  & \multicolumn{2}{c}{\bf mRFP} & \multicolumn{2}{c}{\bf Vaccine degradation}\\
  \cmidrule(lr){2-3}\cmidrule(lr){4-5}
  &\bf GPN-MSA &\bf HyenaDNA &\bf GPN-MSA &\bf HyenaDNA  \\
\midrule
MLP  &  0.330 & 0.439 &  0.572 & 0.695 \\
XGBoost & 0.479 & 0.512 & 0.750 & 0.711 \\

\bottomrule
\end{tabular}
\end{table}

On degradation, XGBoost shows better performance than MLP but still trails specialized RNA models. This consistent gap underscores the need for codon-aware adaptation --- the very role of our CodonMoE module, which injects targeted mRNA context and unlocks DNA backbones’ full predictive power.

\section{Conclusion and discussion} 
\label{conlusion-discussion}
Our CodonMoE is a versatile approach that enhances DNA language models for RNA applications. The framework demonstrates notable adaptability across different model architectures and species, while maintaining strong performance in mRNA-related tasks. The key advantage lies in its ability to leverage the more abundant DNA data for training, while still performing effectively on RNA-specific applications.
This approach represents a significant step toward unifying genomic language modeling by allowing a single model class to handle both DNA and RNA tasks efficiently, reducing computational costs while maintaining or exceeding the performance of specialized RNA models.

\textbf{Model complexity and task requirements}
Performance patterns reveal key relationships between model architecture and biological tasks. Linear complexity models perform well on vaccine stability prediction but show lower gains on mRFP expression. This reflects underlying biological mechanisms: vaccine stability depends primarily on global RNA structure (where state-space models demonstrate superior performance), while mRFP expression requires fine-grained local translation dynamics. The HyenaDNA+CodonMoE combination creates an effective hybrid bridging this local-global gap.

\textbf{Parameter efficiency and the ``threshold effect''}
CodonMoE-pro achieves superior performance with fewer parameters than standard CodonMoE, and HyenaDNA+CodonMoE-pro (7.5M) outperforms configurations with significantly more parameters. This ``threshold effect'' suggests that once an adapter covers critical codon contexts (rare codons, local GC content fluctuations, translation hotspots), additional parameters provide diminishing returns. The optimal range appears to be 7--12M parameters, suggesting that focused biological coverage is more important than parameter count.

\textbf{Biological interpretation and limitations}
While HyenaDNA+CodonMoE-pro performs with notable superiority on vaccine degradation, mRFP expression, and MLOS tasks, its performance on Tc-riboswitches reveals limitations. Riboswitches are non-coding regulatory elements whose function depends on global folding dynamics rather than codon usage patterns. This illuminates a key distinction: CodonMoE demonstrates superior performance at translation-related signals but is less equipped for structural features that dominate riboswitch function. Despite this limitation, it still achieves competitive performance on tasks outside its primary design focus.

\newpage
% \section*{References}
\subsubsection*{Acknowledgments}
We are thankful to Dr.\ Guillaume Mar\c{c}ais  for reviewing this manuscript. This work was supported in part by the US National Science Foundation [III-2232121] and the US National Institutes of Health [R01HG012470].

\subsubsection*{Conflict of Interest}
C.K. is a co-founder of Ocean Genomics, Inc.
\bibliography{codonmoe}

\bibliographystyle{abbrvnat}

\newpage

\appendix

\section{Technical appendices and supplementary material}

\subsection{Algorithm pseudocode}
\label{algorithmpseudocode}
The proposed CodonMoE, whose pseudocode is given in Algorithm~\ref{alg:CodonMoE}, efficiently analyzes mRNA sequences by leveraging a novel MoE model tailored for codon-level feature extraction. This method is designed to operate on the hidden representations produced by a base model trained on DNA sequences, improving mRNA sequence analysis through a codon-level adapter. Below, we outline the core components of this algorithm. 

\paragraph{Input and hidden representation.}

The algorithm takes as input hidden states \( H \in \mathbb{R}^{\text{batch\_size} \times \text{seq\_len} \times d_{\text{model}}} \), where \( H \) is the latent representation generated by a base model trained on nucleotide-level tokenized DNA sequences. These hidden states encapsulate nucleotide-level patterns in the DNA sequence but lack the explicit codon-level representation required for understanding mRNA translation and regulation. CodonMoE restructures these hidden states to focus on codon-level features for better-adapting DNA models for mRNA analysis.

\paragraph{Codon aggregation and reshaping.}

mRNA sequences consist of codons, which are triplets of nucleotides fundamental to protein synthesis. The hidden states \( H \) are reshaped into groups of three consecutive hidden vectors to form codon-level representations. Specifically, the tensor is reshaped into \( [B, S/3, 3d] \), where each codon consists of three concatenated hidden vectors. This step captures interactions between nucleotides within each codon.

\paragraph{Mixture of Experts (MoE) for codon-level feature learning.}

At the core of the CodonMoE is a \textbf{MoE} mechanism that selects from multiple expert networks to process codon-level representations dynamically. Each codon is processed by \( \text{num\_experts} \) linear sub-networks (experts), where each expert specializes in extracting different semantic aspects of the codon. The outputs of these experts are weighted by a softmax gating mechanism, conditioned on the codon input. This ensures the CodonMoE mechanism is highly adaptable to varying contexts within RNA sequences.

\paragraph{Codon-level expansion and integration.}

After extracting codon-level features from the MoE, these features are expanded to match the original sequence length by repeating the codon features three times, once for each nucleotide in the codon. This expanded representation is reshaped back to \( [B, S-1, d] \) and added element-wise to the original hidden states. The result is an enhanced representation that incorporates both nucleotide-level and codon-level information, improving the model’s ability to capture local patterns and broader codon interactions.

\paragraph{Regularization and transformation.}

To ensure robust learning and prevent overfitting, the algorithm applies a series of regularization and transformation steps:
\begin{itemize}
    \item \textbf{Layer normalization}: Ensures stability during training by normalizing the feature map.
    \item \textbf{GELU activation}: Introduces non-linearity to enhance the model's ability to learn complex relationships between codon sequences and biological function.
    \item \textbf{Dropout}: Prevents overfitting by randomly dropping units during training, particularly useful for high-dimensional biological data.
\end{itemize}
The final feature map is then flattened and passed through a linear transformation, producing a compact feature vector \( Y \) that can be used for downstream tasks, such as mRNA classification or regression.

\begin{algorithm}[htbp]
\caption{CodonMoE for mRNA Sequence Analysis}
\label{alg:CodonMoE}
\begin{algorithmic}[1]
  \Require Hidden states $H \in \mathbb{R}^{\text{batch\_size}\times \text{seq\_len}\times d_{\text{model}}}$
  \Ensure Feature vector $Y$
  \Statex
  \State Hyperparameters: $\text{num\_experts}\gets4$, $\text{dropout\_rate}\gets0.1$
  \Statex
  \Function{MixtureOfExperts}{$X$}
    \For{$i\gets1$ \textbf{to} $\text{num\_experts}$}
      \State $\text{expert}_i \gets \text{Sequential}(\text{Linear}(3d,3d),\;\text{GELU},\;\text{Linear}(3d,d))$
      \State $\text{outputs}[i] \gets \text{expert}_i(X)$
    \EndFor
    \State $\text{gate} \gets \mathrm{Softmax}\bigl(\text{Linear}(3d,\text{num\_experts})(X)\bigr)$
    \State \Return $\sum_{i=1}^{\text{num\_experts}} \text{outputs}[i]\;\odot\;\text{gate}[:,:,i]$
  \EndFunction
  \Statex
  \Function{CodonMoE}{$H$}
    \State $(B,S,d)\gets\mathrm{shape}(H)$
    \State $Y\gets H[:,\,0:S-1,\,:]$
    \State $\text{codons}\gets \mathrm{Reshape}\bigl(Y,\,[B,\,\lfloor(S-1)/3\rfloor,\,3d]\bigr)$
    \State $\text{moe}\gets \text{MixtureOfExperts}(\text{codons})$
    \State $\text{expanded}\gets\mathrm{Repeat}(\text{moe},\,3,\,\text{dim}=1)$
    \State $\text{expanded}\gets \mathrm{Reshape}\bigl(\text{expanded},\,[B,\,S-1,\,d]\bigr)$
    \State $Y\gets Y + \text{expanded}$
    \State $Y\gets\mathrm{Dropout}\bigl(\mathrm{GELU}(\mathrm{LayerNorm}(Y)),\,\text{dropout\_rate}\bigr)$
    \State $Y\gets \text{Linear}\bigl((S-1)d,\,d\bigr)\bigl(\mathrm{Flatten}(Y)\bigr)$
    \State $Y\gets\mathrm{Dropout}\bigl(\mathrm{GELU}(\mathrm{LayerNorm}(Y)),\,\text{dropout\_rate}\bigr)$
    \State \Return $\text{Linear}(d,1)(Y)$
  \EndFunction
  \Statex
  \Function{Analyze\_mRNA}{$\text{sequence}$}
    \State $\text{tokens}\gets \mathrm{Tokenize}(\text{sequence})$
    \State $\text{hidden}\gets \mathrm{BaseModel}(\text{tokens})$
    \State \Return \textsc{CodonMoE}(\text{hidden})
  \EndFunction
\end{algorithmic}
\end{algorithm}

\subsection{Proof of Theorem~\ref{thm:universal_approximation}}
\label{proof}

\begin{proof}
We aim to show that the CodonMoE, functioning as an adapter to the pretrained DNA backbone $h$, is a universal approximator for any function $f \in \mathcal{F}$, where $\mathcal{F}$ is the class of continuous functions mapping codon sequences to target properties.

Let $x \in \mathcal{X}$ be a sequence partitioned into $n$ codons:
\[
x = (c_1, c_2, \dots, c_n), \quad c_i \in \mathcal{C}.
\]
The backbone model $h: \mathcal{X} \to \mathbb{R}^{L \times D}$ with $L = 3n$ generates embeddings:
\[
h(x) = [e_1, e_2, \dots, e_L]^\top \in \mathbb{R}^{L \times D}.
\]
   Each codon $c_i$ is represented by averaging three nucleotide embeddings:
   \[
   z_i = \frac{e_{3i-2} + e_{3i-1} + e_{3i}}{3} \in \mathbb{R}^{D}.
   \]

The CodonMoE applies a Mixture of Experts model to each $z_i$:
\[
f_{\text{MoE}}(z_i) = \sum_{k=1}^K g_k(z_i) \cdot E_k(z_i),
\]
where:
\[
g_k(z_i) = \frac{\exp(\phi_k(z_i))}{\sum_{j=1}^K \exp(\phi_j(z_i))},
\]
with gating functions $\phi_k: \mathbb{R}^{D} \to \mathbb{R}$, and expert networks $E_k: \mathbb{R}^{D} \to \mathbb{R}^m$.

By the Universal Approximation Theorem \citep{hornik1989multilayer}, for each $f_k$ and any $\epsilon > 0$, there exists $E_k$ such that:
\[
\| E_k(z_i) - f_k(z_i) \| < \frac{\epsilon}{K n},
\]
where $f_k \in C(\mathbb{R}^{D}, \mathbb{R}^m)$.

Define the overall network function:
\[
F(x) = \sum_{i=1}^n f_{\text{MoE}}(z_i) = \sum_{i=1}^n \sum_{k=1}^K g_k(z_i) \cdot E_k(z_i).
\]
For the target function $f \in \mathcal{F}$:
\[
f(x) = \sum_{i=1}^n f_i(z_i), \quad f_i \in C(\mathbb{R}^{D}, \mathbb{R}^m).
\]
Then, the approximation error is:
\[
\| F(x) - f(x) \| = \left\| \sum_{i=1}^n \sum_{k=1}^K g_k(z_i) \cdot E_k(z_i) - \sum_{i=1}^n f_i(z_i) \right\|.
\]
Given that $\sum_{k=1}^K g_k(z_i) = 1$ and $g_k(z_i) \geq 0$, we have:
\[
\| F(x) - f(x) \| \leq \sum_{i=1}^n \sum_{k=1}^K g_k(z_i) \| E_k(z_i) - f_i(z_i) \| < \sum_{i=1}^n \sum_{k=1}^K g_k(z_i) \frac{\epsilon}{K n}  = \frac{\epsilon}{K}.
\]

The backbone model $h$ ensures that embeddings $z_i$ capture essential genetic information:
\[
h: \mathcal{X} \to \mathbb{R}^{L \times D}, \quad z_i = \mathcal{P}(h(x)),
\]
where $\mathcal{P}$ denotes the partitioning into codon embeddings via averaging.

Combining the above, for any $f \in \mathcal{F}$ and $\epsilon > 0$, there exists a CodonMoE network such that:
\[
\| F(x) - f(x) \| < \epsilon.
\]
Thus, the CodonMoE integrated with the pretrained backbone $h$ satisfies:
\[
F = \sum_{i=1}^n \sum_{k=1}^K g_k(z_i) \cdot E_k(z_i) \approx f(x), \quad \forall f \in \mathcal{F}.
\]

Therefore, the CodonMoE module, when combined with the pretrained backbone model $h$, serves as a universal approximator for any continuous function mapping codon sequences to target properties within the class $\mathcal{F}$.

\end{proof}

\subsection{Additional experimental details}
\label{appendix:experimentalsettings}
\textbf{Experimental settings.} Table~\ref{table7} outlines the key components and hyperparameters used for different backbone models, highlighting the settings in regressor types and training parameters such as learning rates and the number of epochs. Specifically, it details the setup for the mRFP expression dataset, using Caduceus and HyenaDNA as primary backbones with variations such as Caduceus+CodonMean and Caduceus+CodonMoE, indicating different CodonMoE variations within the same framework. Specific configurations such as the backbone sequence length, model dimensions, number of layers, and learning rates are listed, with pure backbone models integrating machine learning regressors like MLP and XGBoost. It also outlines settings for the SARS-CoV-2 vaccine degradation dataset with similar backbone models but slightly adjusted parameters, such as a different sequence length for the HyenaDNA models. This table showcases the learning rates and epochs where applicable, providing a comprehensive view of how each model is tuned for its respective task. 
\begin{table}[htbp]
\centering
\caption{Summary of experimental settings for SARS-CoV-2 vaccine degradation dataset and mRFP expression dataset.}
\resizebox{\textwidth}{!}{
\label{table7}
\begin{tabular}{cccccc}
\toprule
\textbf{Backbone} & \textbf{Model} & \textbf{Backbone Name} & \textbf{Regressor} & \textbf{Learning Rate} & \textbf{Epochs} \\
\midrule
Caduceus & Caduceus & caduceus-ps\_seqlen-1k\_d\_model-256\_n\_layer-4\_lr-8e-3 & mlp & - & - \\
Caduceus & Caduceus & caduceus-ps\_seqlen-1k\_d\_model-256\_n\_layer-4\_lr-8e-3 & xgboost & - & - \\
Caduceus & Caduceus+CodonMean & caduceus-ps\_seqlen-1k\_d\_model-256\_n\_layer-4\_lr-8e-3 & - & 0.0005 & 100 \\
Caduceus & Caduceus+CodonMoE & caduceus-ps\_seqlen-1k\_d\_model-256\_n\_layer-4\_lr-8e-3 & - & 0.0005 & 100 \\
HyenaDNA & HyenaDNA & hyenadna-small-32k-seqlen & mlp & - & - \\
HyenaDNA & HyenaDNA & hyenadna-small-32k-seqlen & xgboost & - & - \\
HyenaDNA & HyenaDNA+CodonMean & hyenadna-small-32k-seqlen & - & 0.0005 & 100 \\
HyenaDNA & HyenaDNA+CodonMoE & hyenadna-small-32k-seqlen & - & 0.0001(0.001) & 100 \\
\bottomrule
\end{tabular}
}
\end{table}

\textbf{Dataset details.} 
For the mRFP expression dataset, \citet{nieuwkoop2023revealing} constructed low (CAI\_L), medium (CAI\_M), and high (CAI\_H) CAI libraries and expressed them in Escherichia coli DH10B. They quantified mRFP expression using both flow cytometry and microplate reader measurements, normalizing fluorescence to account for variations in cell density. The full-length coding sequence (675 bp) for each variant was determined by Sanger sequencing.
They applied quality control criteria to ensure data integrity, excluding samples with low-quality sequencing reads, amino acid mutations, mixed populations, or significant deviations between measurement methods. This curation process resulted in a high-quality dataset that provides a foundation for investigating the determinants of translation efficiency in \textit{E. coli}.
We accessed this dataset through the public repository as provided by the original authors and used it as the basis for our machine learning approach to predict protein production levels from mRNA sequence features.

For the SARS-Cov-2 vaccine degradation dataset, this dataset includes mRNA constructs encoding a multi-epitope vaccine (MEV) candidate based on SARS-CoV-2 antigens. The key component of this dataset that we focus on in our experiments is the in-cell mRNA stability via time-course degradation experiments in HEK293T cells. This dataset, as described by~\citet{leppek2022combinatorial}, provides a resource for investigating the relationships between mRNA sequence, structure, stability, and expression efficiency in the context of SARS-CoV-2 vaccine design.

The Tc-riboswitch dataset~\citep{groher2018tuning} was developed to optimize the dynamic range (DR) and basal expression (BE) of tetracycline (Tc)-responsive synthetic riboswitches. These constructs consist of tandem Tc aptamers inserted into the $5^{\prime}$ untranslated region (UTR) of a GFP reporter gene, regulating expression in response to Tc ligand binding. Using \textit{Saccharomyces cerevisiae} RS453 as the host, GFP fluorescence was quantified with and without Tc induction via flow cytometry. Through machine learning-guided optimization, including random forest classifiers and convolutional neural networks, sequence and structural features influencing DR and BE were systematically explored by~\citet{groher2018tuning}. The curated dataset includes constructs with optimized biophysical properties, providing a foundation for understanding riboswitch function and advancing ML-driven design frameworks.

The MLOS dataset~\citep{li2024codonbert} contains 164 mRNA candidates designed to encode the influenza hemagglutinin antigen, constructed with fixed untranslated regions and variable coding regions. These candidates were synthesized and transfected into cells to evaluate their performance. Additionally, the benchmarking dataset incorporated sequences from Sanofi encoding the hemagglutinin antigen for flu vaccines. Specifically, these mRNA sequences were evaluated for protein expression levels in HeLa cells.

\subsection{Feature embedding visualization}
\label{appendix:visualization}
\begin{figure}[ht]
\centering\includegraphics[width=0.78\textwidth]{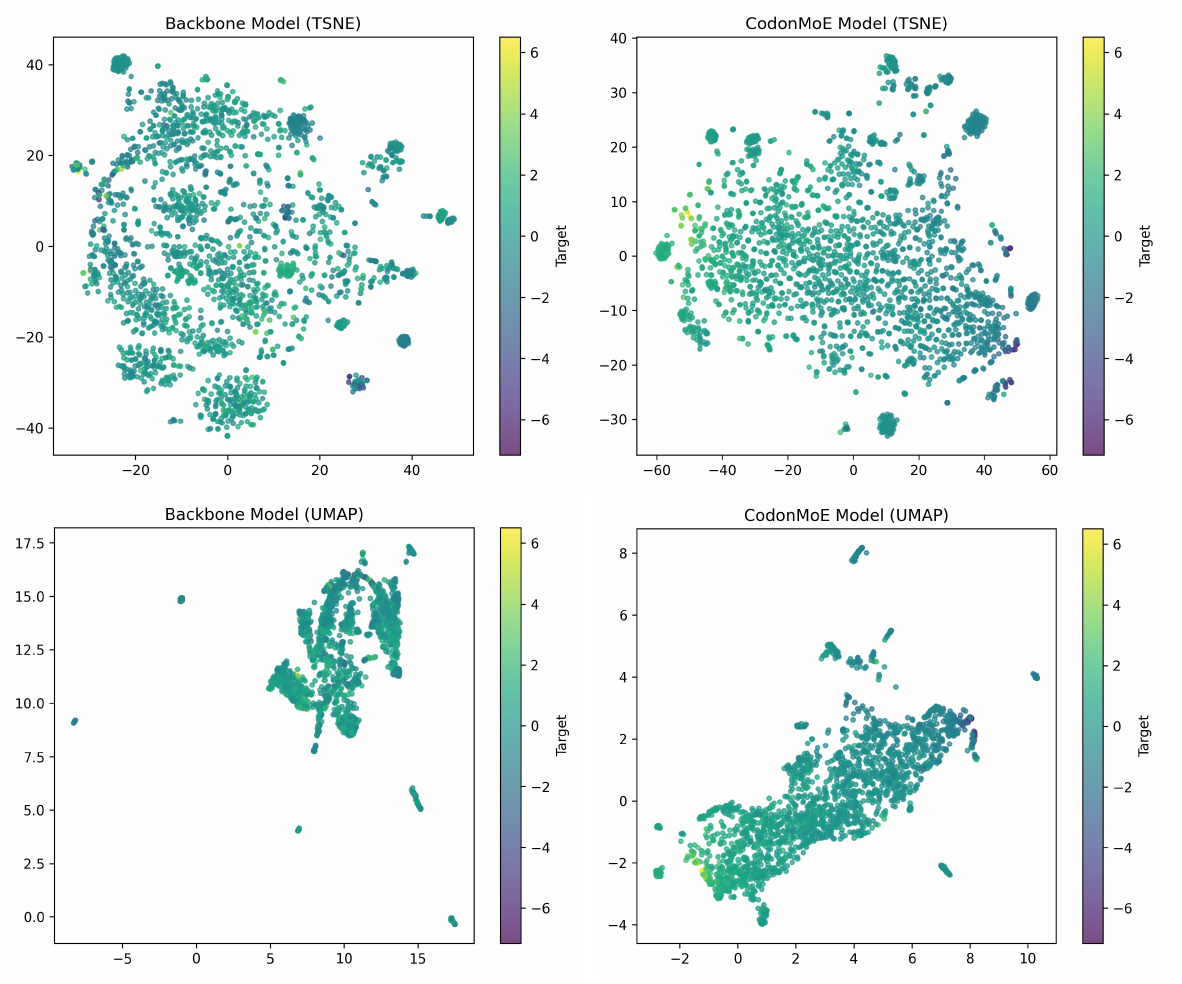}
    \caption{t-SNE and UMAP comparison between features from HyenaDNA model and CodonMoE-enhanced HyenaDNA model on SARS-CoV-2 vaccine degradation dataset.}
    \label{fig:hyena_cov_vis}
\end{figure}

\textbf{SARS-CoV-2 vaccine degradation task.} As shown in Figure~\ref{fig:hyena_cov_vis}, the UMAP and t-SNE visualizations highlight the CodonMoE model's superior ability to capture fine-grained codon-level patterns and dynamically specialize through its Adaptive Mixture of Experts, resulting in more distinct and diverse clusters compared to the backbone model. CodonMoE’s expert system allows for better separation of genetic features, capturing both local codon-specific and broader sequence patterns. This leads to smoother transitions in the continuous target values, as seen in the clearer color gradients in the t-SNE plot, indicating that CodonMoE is able to approximate complex relationships between codon sequences and degradation rates. In contrast, the backbone model’s visualizations show more compressed clusters and limited separation, suggesting that it struggles with representing nuanced degradation patterns. 

\begin{figure}[ht]
    \centering
    \includegraphics[width=0.78\textwidth]{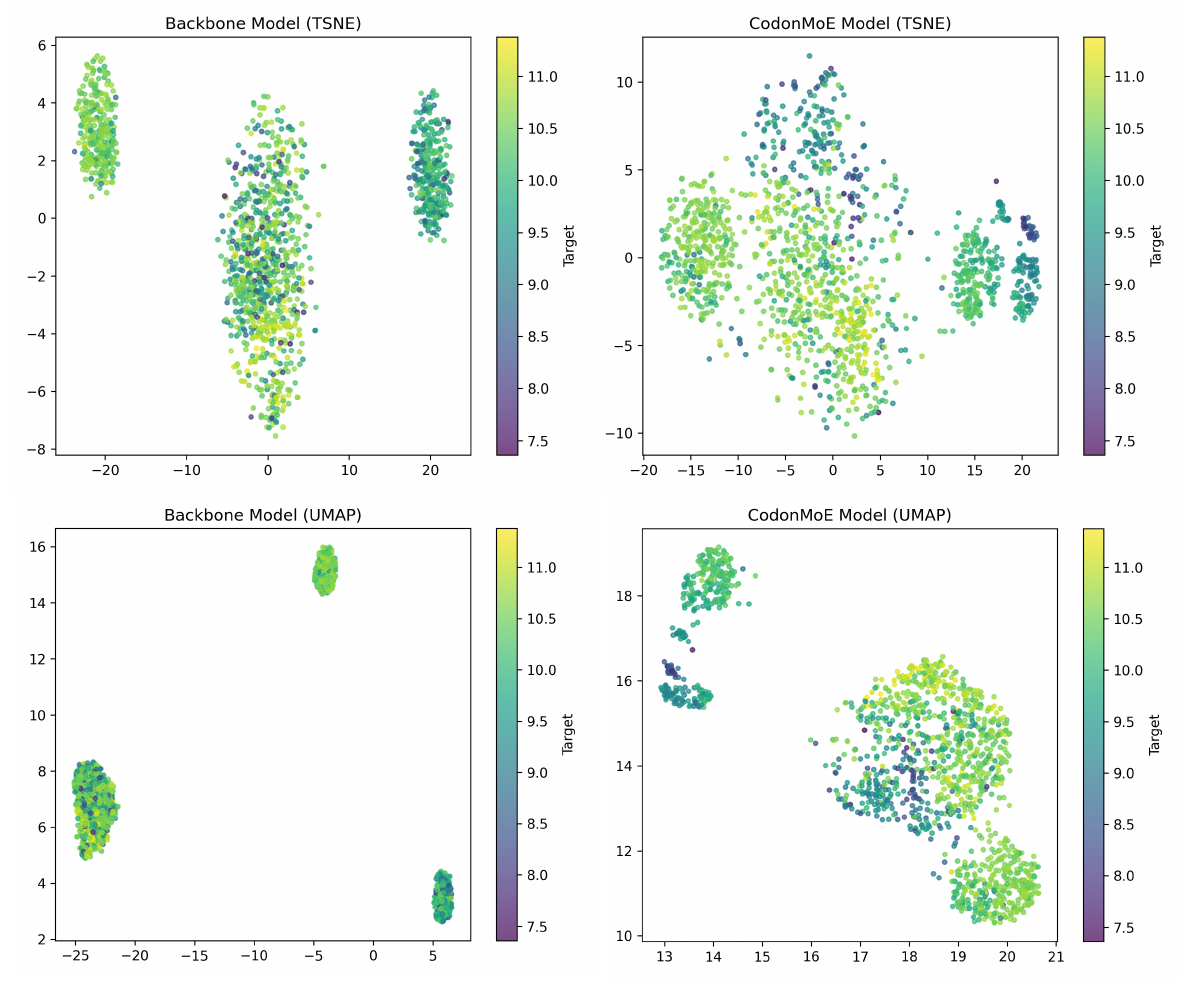}
    \caption{t-SNE and UMAP comparison between features from HyenaDNA model and CodonMoE-enhanced HyenaDNA model on mRFP expression dataset.}
    \label{fig:hyena_mrfp_vis}
\end{figure}

\begin{figure}[ht]
    \centering
    \includegraphics[width=0.78\textwidth]{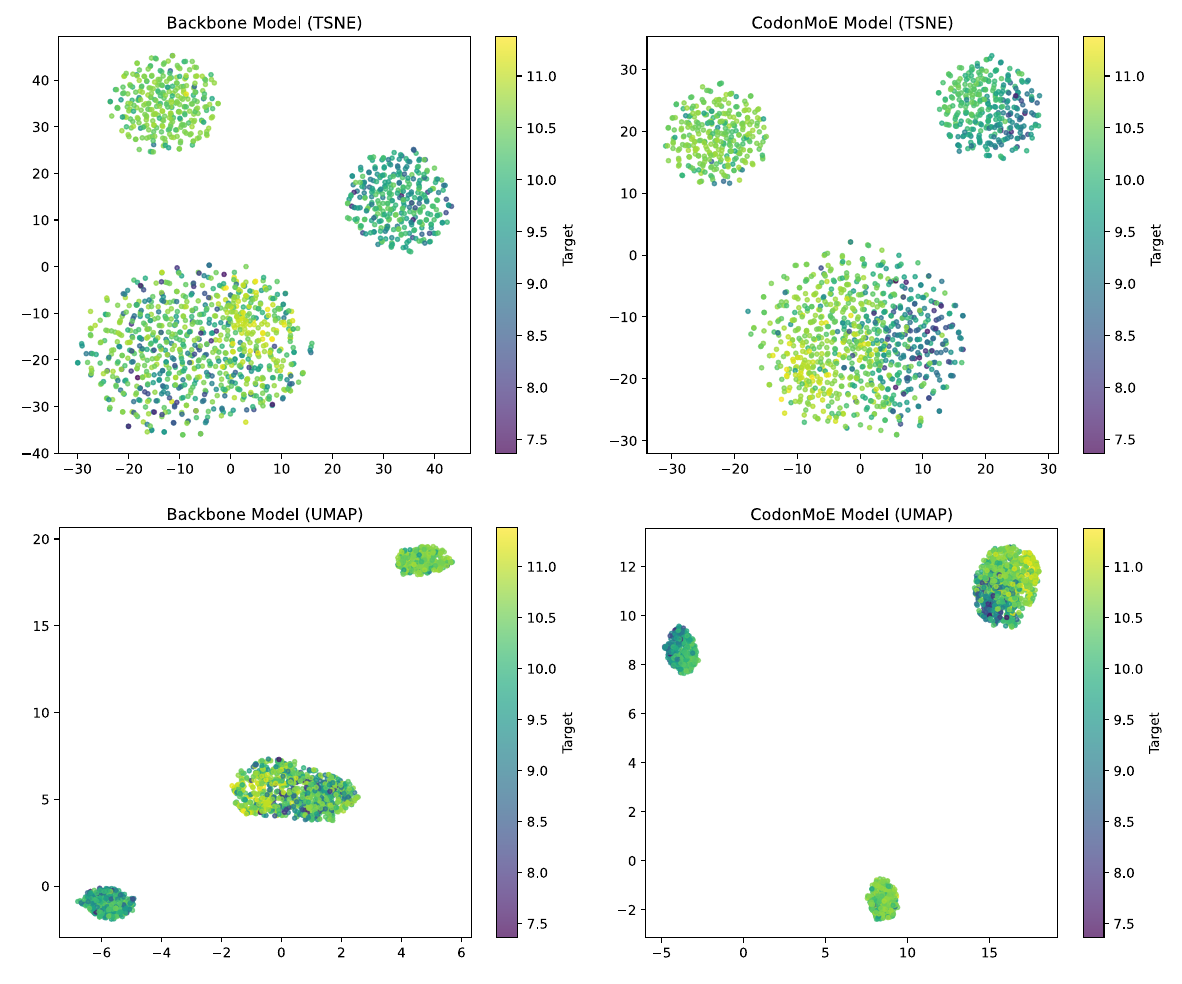}
    \caption{t-SNE and UMAP comparison between features from GPN-MSA model and CodonMoE-enhanced GPN-MSA model on mRFP expression dataset.}
    \label{fig:gpn_mrfp_vis}
\end{figure}

\textbf{mRFP expression task.} In Figure~\ref{fig:hyena_mrfp_vis}, the t-SNE and UMAP visualizations highlight the improved performance of the CodonMoE-enhanced HyenaDNA model compared to the backbone model on the mRFP expression dataset. In the t-SNE plot, the backbone model shows tight clusters with limited spread, indicating that it struggles to differentiate between various expression levels, leading to more uniform representations. In contrast, CodonMoE demonstrates broader, more distinct clusters, reflecting its ability to capture finer differences in mRFP expression levels, as seen in the smoother color gradient transitions. Similarly, the UMAP visualization reveals that the backbone model's clusters are tightly packed, suggesting less feature diversity, whereas CodonMoE’s clusters are more spread out, indicating richer, more nuanced representations. This enhanced separation and feature diversity in CodonMoE can be attributed to its architecture, which allows it to capture both local codon-level patterns and broader sequence features, resulting in better predictions of continuous targets like mRFP expression levels.
Figure~\ref{fig:gpn_mrfp_vis} shows that the CodonMoE-enhanced GPN-MSA model demonstrates clearer and more distinct clustering. In both t-SNE and UMAP visualizations, the CodonMoE-enhanced backbone features tighter and more defined clusters with a pronounced variation in metric values, suggesting a more effective differentiation.

\subsection{Additional experiments on evaluation of DNA pretrained model feature effectiveness}
\label{appendix：codonmoepro_effectiveness}
\begin{figure}[htbp]
    \centering
    \begin{minipage}[t]{0.45\textwidth}
        \centering
        \includegraphics[width=\textwidth]{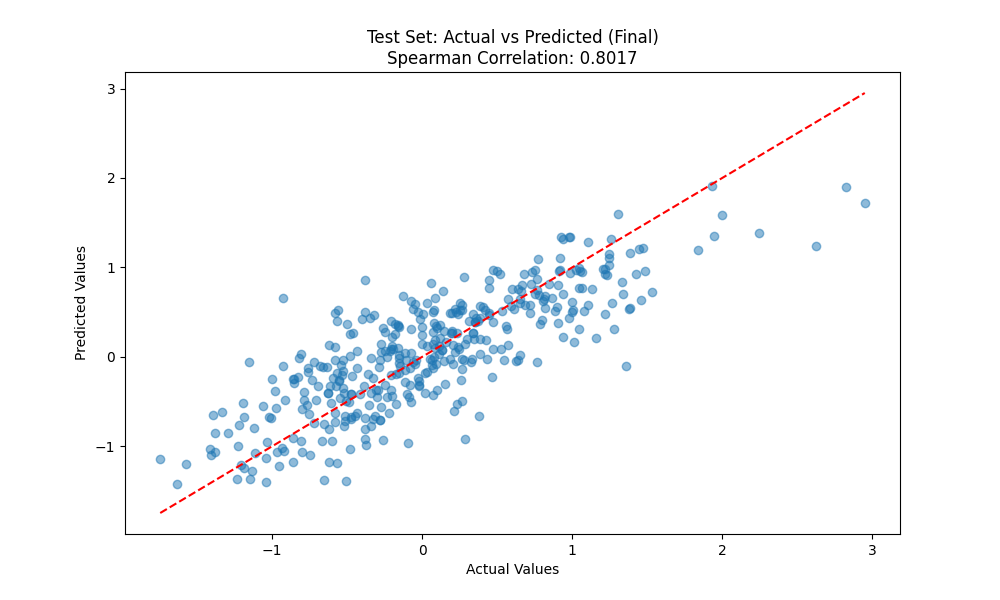}
        \vspace{0.5em}
        \\(a) Caduceus pretrained model feature effectiveness on SARS-CoV-2 vaccine degradation dataset.
        \label{fig:tconvolcodon_Caduceus_Cov}
    \end{minipage}
    \hfill
    \begin{minipage}[t]{0.45\textwidth}
        \centering
        \includegraphics[width=\textwidth]{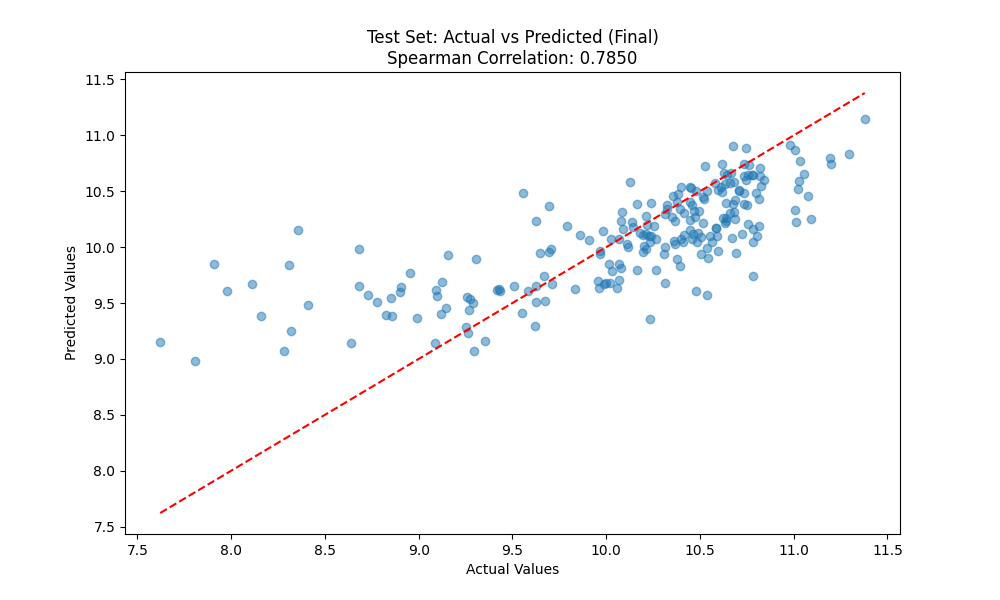}
        \vspace{0.5em}
        \\(b) Caduceus pretrained model feature effectiveness on mRFP expression dataset.
        \label{fig:tconvolcodon_Caduceus_mRFP}
    \end{minipage}
    
    \vspace{1em}
    
    \begin{minipage}[t]{0.45\textwidth}
        \centering
        \includegraphics[width=\textwidth]{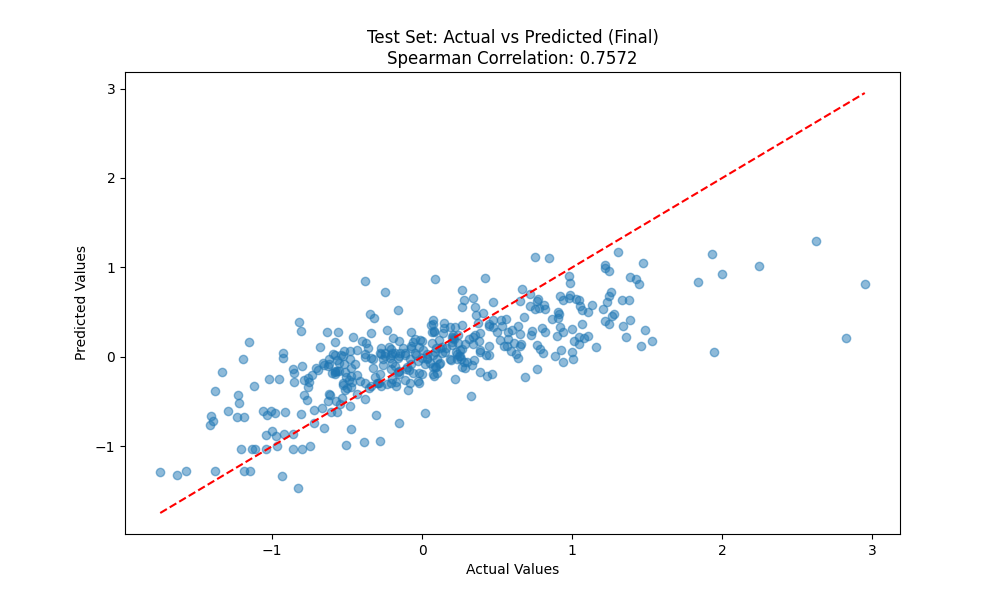}
        \vspace{0.5em}
        \\(c) HyenaDNA pretrained model feature effectiveness on SARS-CoV-2 vaccine degradation dataset.
        \label{fig:tconvolcodon_HyenaDNA_Cov}
    \end{minipage}
    \hfill
    \begin{minipage}[t]{0.45\textwidth}
        \centering
        \includegraphics[width=\textwidth]{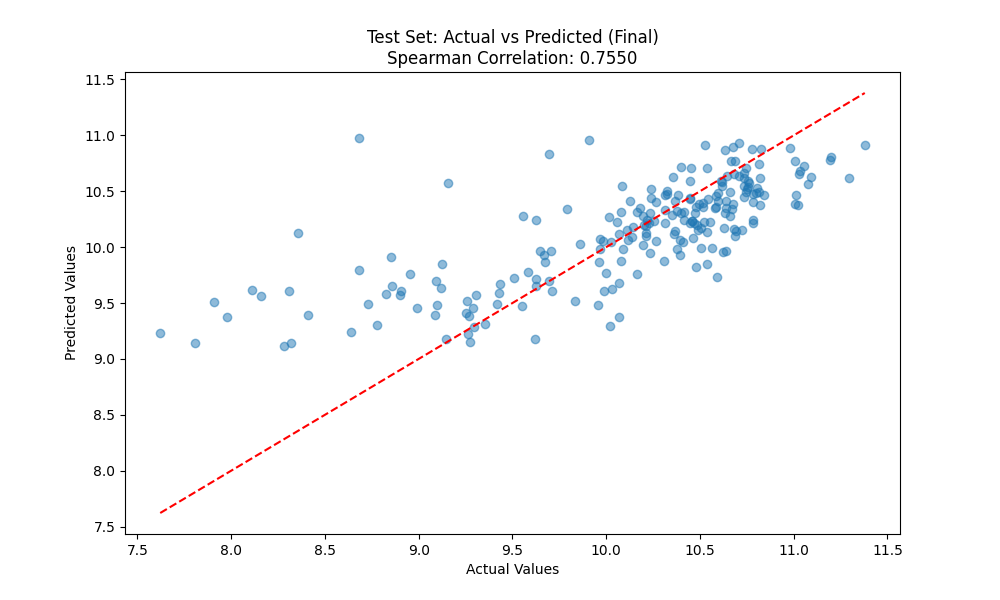}
        \vspace{0.5em}
        \\(d) HyenaDNA pretrained model feature effectiveness on mRFP expression dataset.
        \label{fig:tconvolcodon_HyenaDNA_mRFP}
    \end{minipage}
    
    \caption{Evaluation of DNA pretrained model feature effectiveness on mRFP expression and SARS-CoV-2 vaccine degradation dataset using TextCNN.}
    \label{fig:tconvolcodon_Comparison}
\end{figure}

In this section, we explore the ability of DNA-pretrained backbones, specifically Caduceus and HyenaDNA, to effectively generalize to mRNA-related tasks using a TextCNN framework. The tasks evaluated include predictions on the SARS-CoV-2 vaccine degradation dataset and the mRFP expression dataset. The scatter plots in Figure~\ref{fig:tconvolcodon_Comparison} provide a visual representation of the alignment between actual and predicted values, with a trendline indicating overall correlation.

The SARS-CoV-2 vaccine degradation dataset serves as a proxy for evaluating the potential of DNA-pretrained features to capture complex biological dependencies related to RNA sequence stability and degradation. Both Caduceus and HyenaDNA demonstrate a clear trend of alignment between actual and predicted values, reflecting the potential of DNA-derived features to transfer effectively to mRNA stability prediction. While the inherent challenges of modeling degradation, as indicated by a wider spread in predictions, the performance reflects the potential of pretrained DNA models to generalize beyond their training domain to tasks with overlapping biological mechanisms, such as RNA stability. The effectiveness of these features suggests that key structural and sequence-specific attributes learned from DNA datasets are applicable to mRNA-related degradation tasks.

The mRFP expression dataset focuses on the predictability of gene expression levels based on underlying sequence features. Both models achieve a closer alignment of predicted values to the actual values compared to the degradation dataset. This suggests that the DNA-pretrained features can be potentially effective at tasks involving expression prediction, where sequence features such as promoter regions, codon optimization, and untranslated regions are critical. The high clustering around the trendline demonstrates that these DNA backbones successfully capture sequence motifs and structural patterns that are transferable to mRNA-related tasks. This finding aligns with the hypothesis that DNA and RNA share significant overlapping biological motifs, enabling effective transfer learning.

\subsection{Additional ablation studies}
\label{appendix：ablation_dense}
To further evaluate the effectiveness of the CodonMoE architecture, we conducted additional experiments comparing its performance with a dense baseline model. The results are summarized in Table~\ref{tab:dense_ablation}. The dense baseline replaces the CodonMoE module with standard dense layers while maintaining an equivalent number of trainable parameters and identical training hyperparameters, ensuring a controlled setup for fair ablation studies. This approach isolates the contribution of the CodonMoE architecture to the overall performance.

The consistent performance gains across both datasets indicate that CodonMoE's specialized design provides superior modeling capabilities compared to standard dense layers under matched parameter constraints. This reinforces the potential of CodonMoE as a plug-and-play module for adapting DNA-based models to mRNA tasks, offering both computational efficiency and improved predictive performance.

\begin{table}[htbp]
\centering
\caption{Performance comparison between the standard dense baseline and HyenaDNA+CodonMoE-pro (equivalent parameters) on SARS-CoV-2 vaccine degradation dataset and RFP expression dataset.}
\label{tab:dense_ablation}
% \color{blue}
\begin{tabular}{lcc}
\hline
Model & Vaccine Degradation & mRFP Expression \\ \hline
HyenaDNA-Densebaseline  & 0.80 & 0.82 \\
HyenaDNA+CodonMoE-pro& \textbf{0.84} & \textbf{0.88} \\ \hline
\end{tabular}
\end{table}

\subsection{Additional introduction of DNA backbones}
\label{appendix: intro2backbone}

\subsubsection{RNABERT}
RNABERT is a nucleotide-based RNA large language model trained on non-coding RNAs (ncRNAs) to provide effective embeddings of RNA bases. It integrates context-sensitive nucleotide information with secondary structural features to enhance its understanding of RNA functionality. Trained on 76,237 non-coding RNA sequences from RNAcentral using masked language modeling and structural alignment learning, RNABERT excels in capturing both nucleotide-level interactions and higher-order structural similarities that underpin RNA functionality. The architecture of RNABERT, comprising 6 Transformer layers with a hidden dimension of 120. 

\subsubsection{RNA-FM}
RNA-FM is a nucleotide-based foundational RNA language model specifically designed for large-scale RNA structure and function prediction. RNA-FM employs a 12-layer bidirectional Transformer encoder to capture intricate long-range interactions and evolutionary signals within RNA sequences. Trained on 23 million unannotated ncRNA sequences from RNAcentral using self-supervised learning, RNA-FM generates highly expressive embeddings that represent both structural and functional characteristics. Despite its larger architecture, RNA-FM demonstrates high efficiency, offering robust generalization across diverse RNA datasets while requiring less fine-tuning for new tasks. Its flexibility and precision make RNA-FM a cornerstone model for advancing RNA research across multiple domains.

\subsubsection{CodonBERT}
CodonBERT is a codon-based RNA language model built on the BERT architecture, featuring a 12-layer bidirectional Transformer encoder with 12 self-attention heads per layer and a hidden dimension of 768 at each position. It is pre-trained on 10 million mRNA coding sequences (CDS) sourced from NCBI, covering mammals, bacteria, and human viruses across 13 evolutionary categories. Input sequences are split into codons (triplets of nucleotides) and encoded through a combination of codon embeddings, positional embeddings, and segment embeddings, resulting in context-aware codon representations for downstream tasks. In addition to the Masked Language Modeling (MLM) task, CodonBERT incorporates Homologous Sequence Prediction (HSP), where pairs of mRNA sequences are classified to determine their evolutionary relationships, aiding in the learning of sequence homology. The sequences are preprocessed to ensure lengths are multiples of three, beginning with the start codon (AUG) and ending with stop codons (UAA, UAG, or UGA). Compared to RNABERT and RNA-FM, which focus on nucleotide-based embeddings and non-coding RNA, CodonBERT leverages codon-level inputs, providing a deeper understanding of translation-related features and evolutionary information, making it particularly effective for tasks like mRNA optimization and protein expression prediction.

\subsubsection{GPN-MSA}

GPN-MSA is a DNA language model optimized for genome-wide variant effect prediction, utilizing a multiple-sequence alignment (MSA) of 100 vertebrate species. These alignment blocks are then stitched together using the multiz utility maf2fasta, ensuring that any columns with gaps in the human reference are removed, and excluding the 10 primate species closest to humans to avoid bias from excessive similarity. Additionally, associated conservation scores from phastCons and phyloP, which provide important information about evolutionary conservation across species, are downloaded and integrated into the training data.

The GPN-MSA model architecture leverages masked language modeling techniques, using a 128-bp multiple-sequence alignment (MSA) window. In this setup, 15\% of the positions within the human reference sequence are masked randomly during training, and the model learns to predict these nucleotides based on the contextual information provided by both the positions and species represented in the MSA. The sequence of MSA columns is processed through a Transformer neural network named RoFormer \citep{su2024roformer}, which results in a high-dimensional contextual embedding for each position, and a final layer outputs the probabilities for four nucleotides at each masked position.

To optimize the learning process, the model downweights repetitive elements and upweights conserved elements, ensuring that incorrect predictions in neutral regions are penalized less severely. A smoothed version of phastCons, referred to as phastConsM, is used to emphasize highly conserved regions and those immediately adjacent to them. As part of data augmentation in non-conserved regions, the reference nucleotide is replaced by a random nucleotide with a certain probability, guiding the model to assign more neutral scores in these less conserved areas. This strategic integration of evolutionary conservation and species diversity, along with sophisticated neural modeling techniques, allows GPN-MSA to effectively learn from a rich and complex set of genomic data, making it a powerful tool for predicting variant effects across the genome.
\subsubsection{HyenaDNA}
HyenaDNA is a genomic foundation model that addresses the challenges of long-range dependencies and single-nucleotide resolution in DNA sequence analysis. Unlike traditional Transformer-based approaches constrained by the quadratic scaling of attention mechanisms, HyenaDNA employs the Hyena operator, which enables ultralong context lengths of up to 1 million tokens. This represents a 500x improvement in context length over previous dense-attention genomic models. Pretrained on the human reference genome using next-nucleotide prediction, HyenaDNA excels in capturing both the intricate long-range interactions within genomic sequences and the subtle single-nucleotide variations that drive biological functions. Its architecture is highly efficient, scaling sub-quadratically in sequence length and training up to 160x faster than Transformers for similar tasks. Despite using significantly fewer parameters and less pretraining data, HyenaDNA achieves state-of-the-art performance across 20+ genomic benchmarks, including enhancer identification and chromatin profile prediction. Moreover, its innovative use of soft prompting and in-context learning allows for rapid adaptation to new genomic tasks without fine-tuning model weights, showcasing its flexibility and broad utility in genomic research.
\subsubsection{Caduceus}
Caduceus is a DNA language model that combines novel architectural innovations to address critical challenges in genomic sequence modeling, including long-range dependencies, bi-directionality, and reverse complement (RC) equivariance. Unlike traditional genomic models, Caduceus leverages the MambaDNA block, a powerful extension of the Mamba module, to process sequences bi-directionally while incorporating RC-equivariant processing as an inductive bias. This ensures that predictions remain invariant under strand reversal, a critical requirement for accurate DNA sequence modeling.

Pretrained on the human reference genome with a masked language modeling (MLM) objective, Caduceus is specifically designed to handle sequences extending to hundreds of thousands of nucleotides, surpassing the limitations of unidirectional models or those reliant on quadratic scaling attention mechanisms. Its RC-equivariant embeddings and prediction heads enhance its ability to capture the symmetry of DNA, making it particularly effective in tasks involving regulatory annotations, enhancer prediction, and variant effect analysis.

The model achieves exceptional performance across a broad range of genomic tasks, including variant effect prediction and enhancer classification, often outperforming significantly larger models such as Nucleotide Transformer v2~\citep{dalla2023nucleotide} and other Transformer-based architectures.

\subsection{Additional model comparison}
\label{appendix:comprehensive_comparison}

To provide a more comprehensive evaluation of our approach against the broader landscape of RNA analysis methods~\citep{li2024codonbert}, Table~\ref{tab:model_comparison} presents an expanded comparison that includes traditional feature-based methods not discussed in the main text. These methods represent important baselines that have historically been used for genomic sequence analysis before the advent of deep learning approaches.

The traditional feature-based approaches include TF-IDF (Term Frequency-Inverse Document Frequency), which treats codons as vocabulary terms and measures their frequency distribution across sequences. Despite its simplicity and lack of trainable parameters, TF-IDF achieves reasonable performance on stability prediction tasks on the vaccine degradation dataset. 

Plain TextCNN applies convolutional neural networks directly to codon sequences without extensive pre-training, using shallow architectures with approximately 0.15M parameters. This lightweight approach performs surprisingly well on certain tasks, achieving competitive performance on vaccine degradation prediction. 

Codon2vec+TextCNN employs pre-trained codon embeddings (similar to word2vec in natural language processing) before applying convolutional layers, providing stronger semantic representations of codons with minimal parameter overhead. This method shows balanced performance across tasks, particularly on the Tc-riboswitch dataset where it matches several larger models.

The more comprehensive comparison on three datasets demonstrates that while our CodonMoE-enhanced DNA models achieve state-of-the-art performance across most tasks, certain traditional approaches remain competitive for specific applications despite their significantly lower computational requirements. This suggests that task-specific inductive biases can sometimes compensate for model scale, particularly when biological mechanisms align well with the model architecture.

\begin{table}[ht]
\caption{Additional evaluation of computational complexity and Spearman's rank correlation metrics across RNA and DNA language models: measuring the impact of CodonMoE integration on model performance and parameter efficiency. Each data set is split into training, validation, and testing with a 0.7, 0.15, and 0.15 ratio, using the same split set as in the CodonBERT~\citep{li2024codonbert}. The metric is Spearman's rank correlation.}
\label{tab:model_comparison}
\centering
\footnotesize 
\setlength{\tabcolsep}{2.5pt}
\begin{tabular}{lccccccc}
\toprule
\rowcolor{gray!10}
\textbf{Model} & \textbf{Type} & \textbf{Params} & \textbf{Size} & \textbf{Complexity} & \textbf{Vaccine} & \textbf{mRFP} & \textbf{Tc-ribo.} \\
\midrule
\multicolumn{8}{l}{\textit{\textbf{Traditional Feature-based Methods}}} \\
TF-IDF & Codon & -- & -- & $O(N)$ & 0.69 & 0.57 & 0.49 \\
Plain TextCNN & Codon & 0.15M & \textcolor{green}{$\square$} & $O(L)$ & 0.80& 0.78 & 0.43 \\
Codon2vec+TextCNN & Codon & 0.15M & \textcolor{green}{$\square$} & $O(L)$ & 0.70 & 0.77 & 0.56 \\

\midrule
\multicolumn{8}{l}{\textit{\textbf{DNA Foundation Models Enhanced with CodonMoE}}} \\
\rowcolor{blue!10}
HyenaDNA + CodonMoE-pro (ours)& Codon & 7.50M & \textcolor{green}{$\square$} & $O(L\log L)$ & \textbf{0.84} & \textbf{0.88} & \textbf{0.60}\\
\bottomrule
\end{tabular}
\vspace{1mm}
\begin{tablenotes}
\footnotesize
\item \textbf{Notes:} \textbf{Type:} Codon (codon-based), Nucl. (nucleotide-based); \textbf{Size:} \textcolor{green}{$\square$} < 10M, \textcolor{orange}{$\square\square$} = 20-80M, \textcolor{red}{$\square\square\square$} > 80M; \textbf{Complexity:} $O(L)$ - linear, $O(L\log L)$ - linearithmic, $O(L^2)$ - quadratic, $O(N)$ - linear ($L$ = sequence length, $N$ = vocabulary size); 
\end{tablenotes}
\end{table}

\subsection{Detailed analysis of codon operator variants}
\label{appendix:detailed_ablation}
To uncover \emph{why} certain adapter designs perform better --- and how their inductive biases align with biological signals --- we compare three codon operators and anchor our discussion with diagnostic results and theoretical insights.

\paragraph{CodonMean}  
A minimal adapter that computes the arithmetic mean of the three nucleotide embeddings forming each codon. By aggregating global codon usage bias, CodonMean provides a crude correction to DNA backbones’ neglect of synonymous‐codon effects. On both mRFP expression and vaccine degradation tasks, CodonMean delivers a solid lift in Spearman’s $\rho$ (Table~\ref{table2_mrfp}), demonstrating that even a simple global bias adjustment can unlock substantial gains. However, its uniform averaging inherently masks low‐frequency “hotspot” codons and fails to differentiate context‐specific patterns, which caps its maximum benefit.
\paragraph{CodonMoE}  
To address CodonMean’s blind spots, CodonMoE introduces a small Mixture-of-Experts layer per codon position. Each expert specializes on a subset of codons (e.g., rare codons, GC-rich codons), ensuring that sparsely occurring but functionally critical codons receive dedicated capacity. This expert‐driven routing amplifies rare‐codon signals, yielding an additional improvement over CodonMean (Table~\ref{table2_mrfp}). In effect, CodonMoE acts like a “biological microscope,” correcting the DNA backbone’s insensitivity to low-frequency codon patterns that drive ribosomal pausing and local folding events.
\paragraph{CodonMoE-pro}  
Building on CodonMoE, CodonMoE-pro replaces the final linear projections with a \emph{codon neighborhood convolution}: a narrow sliding-window convolution over adjacent codons. This convolutional layer detects recurring codon pairs or triplets—short motifs known to modulate translation kinetics and mRNA stability. By explicitly encoding these local interactions, CodonMoE-pro delivers a further boost on mRFP expression and vaccine degradation, achieving state-of-the-art accuracy with only 7.5M parameters (Table~\ref{table2_mrfp}). The convolutional filter both denoises irrelevant patterns and amplifies the most discriminative codon motifs, functioning as a specialized “motif lens” that zeroes in on biologically meaningful n-gram signals.

\paragraph{Beyond parameter count}  
Table~\ref{tab:dense_ablation} shows that, even when controlling for parameter count, HyenaDNA+CodonMoE-pro significantly outperforms a standard dense baseline. This gap reveals:

\begin{itemize}
  \item \textbf{Structure over Scale:}  Both models possess the same number of parameters, yet the CodonMoE-pro achieves better performance. This demonstrates that \emph{how} parameters are arranged (expert routing) can be more critical than \emph{how many} parameters exist.

 \item \textbf{Implicit Prior for Codon Heterogeneity:}  
    The gating mechanism encodes an inductive bias that codon contexts vary in importance—MoE can “switch on” experts for rare or functionally critical codons, a nuance that uniform dense layers cannot replicate despite equal capacity.
  \item \textbf{Built-in Regularization through Sparsity:} The sparse gating mechanism in CodonMoE-pro inherently regularizes training by limiting expert updates to relevant codons, reducing overfitting compared to the dense baseline’s all‐to‐all parameter updates.

\end{itemize}

These observations suggest that future mRNA prediction models can prioritize biologically informed inductive biases—such as codon n-gram detection—over brute‐force increases in network size, even when computational budgets allow for larger dense architectures.

In addition, our ablations reveal task-specific biases: stability tasks rely on global and mid-range patterns, whereas expression tasks demand fine-grained n-gram detection, justifying the convolutional inductive bias.

In conclusion, structured design-space exploration explain \emph{why} more expressive adapters—far beyond trivial parameter increases—are architecturally aligned with the underlying biology of translation and mRNA stability. Moreover, the feature visualization comparisons between the backbones with and without CodonMoE align closely with the results presented in Figure~\ref{fig:performance_charts}. For a more detailed discussion of these visualization comparisons and more experiments, refer to Appendix~\ref{appendix:visualization} and \ref{appendix：codonmoepro_effectiveness}.

\end{document}